\documentclass[twoside,twocolumn,9pt]{article}
\usepackage{extsizes}
\usepackage[super,sort&compress,comma]{natbib} 
\usepackage[version=3]{mhchem}
\usepackage[left=1.5cm, right=1.5cm, top=1.785cm, bottom=2.0cm]{geometry}
\usepackage{balance}
\usepackage{mathptmx}
\usepackage{sectsty}
\usepackage{graphicx} 
\usepackage{lastpage}
\usepackage{cancel}
\usepackage[format=plain,justification=justified,singlelinecheck=false,font={stretch=1.125,small,sf},labelfont=bf,labelsep=space]{caption}
\usepackage{float}
\usepackage{fancyhdr}
\usepackage{fnpos}
\usepackage[english]{babel}
\addto{\captionsenglish}{%
  
}
\usepackage{array}
\usepackage{droidsans}
\usepackage{charter}
\usepackage[T1]{fontenc}
\usepackage[usenames,dvipsnames]{xcolor}
\usepackage{setspace}
\usepackage[compact]{titlesec}
\usepackage{hyperref}

\usepackage{epstopdf}

\definecolor{cream}{RGB}{222,217,201}

\usepackage{afterpage}
\usepackage{longtable}
\usepackage[english]{babel}
\usepackage{dcolumn}
\usepackage{bm}
\usepackage{pifont}
\usepackage{amsfonts}
\usepackage{amsmath}
\usepackage{amssymb}
\usepackage{wasysym}
\usepackage{amsbsy}
\usepackage{psfrag}
\usepackage{lineno}
\usepackage{color}
\usepackage{multirow}
\usepackage{cases}
\usepackage{stackengine}
\usepackage{blkarray}
\usepackage{cancel}
\usepackage{mathrsfs}
\usepackage{empheq}
\usepackage{tikz}
\usetikzlibrary{arrows}
\usetikzlibrary{decorations.pathmorphing}
\usepackage{pgfplots}
\usepackage{psfrag}
\usepackage{subfig}
\usepackage{wrapfig}
\usepackage[utf8]{inputenc}
\graphicspath{{./figures/}}

\providecommand\bnabla{\boldsymbol{\nabla}}
\providecommand\bcdot{\boldsymbol{\cdot}}
\newcommand{\boldm}[1]{\boldsymbol{#1}}

\newcommand{\sech}{\mathrm{sech} \,}
\newcommand{\csch}{\mathrm{csch} \,}

\DeclareMathAlphabet{\mathbfsf}{\encodingdefault}{\sfdefault}{bx}{sl}

\newcommand{\Ca}{\mbox{\textit{Ca}}} 

\begin{document}
\pagestyle{fancy}
\thispagestyle{plain}
\fancypagestyle{plain}{
\renewcommand{\headrulewidth}{0pt}
}

\makeFNbottom
\makeatletter
\renewcommand\LARGE{\@setfontsize\LARGE{15pt}{17}}
\renewcommand\Large{\@setfontsize\Large{12pt}{14}}
\renewcommand\large{\@setfontsize\large{10pt}{12}}
\renewcommand\footnotesize{\@setfontsize\footnotesize{7pt}{10}}
\makeatother

\renewcommand{\thefootnote}{\fnsymbol{footnote}}
\renewcommand\footnoterule{\vspace*{1pt}%
\color{cream}\hrule width 3.5in height 0.4pt \color{black}\vspace*{5pt}} 
\setcounter{secnumdepth}{5}

\makeatletter 
\renewcommand\@biblabel[1]{#1}            
\renewcommand\@makefntext[1]%
{\noindent\makebox[0pt][r]{\@thefnmark\,}#1}
\makeatother 
\renewcommand{\figurename}{\small{Fig.}~}
\sectionfont{\sffamily\Large}
\subsectionfont{\normalsize}
\subsubsectionfont{\bf}
\setstretch{1.125} 
\setlength{\skip\footins}{0.8cm}
\setlength{\footnotesep}{0.25cm}
\setlength{\jot}{10pt}
\titlespacing*{\section}{0pt}{4pt}{4pt}
\titlespacing*{\subsection}{0pt}{15pt}{1pt}

\fancyfoot{}
\fancyfoot[LO,RE]{\vspace{-7.1pt}\includegraphics[height=9pt]{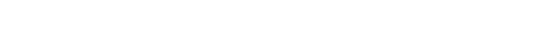}}
\fancyfoot[CO]{\vspace{-7.1pt}\hspace{13.2cm}\includegraphics{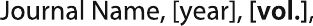}}
\fancyfoot[CE]{\vspace{-7.2pt}\hspace{-14.2cm}\includegraphics{head_foot/RF}}
\fancyfoot[RO]{\footnotesize{\sffamily{1--\pageref{LastPage} ~\textbar  \hspace{2pt}\thepage}}}
\fancyfoot[LE]{\footnotesize{\sffamily{\thepage~\textbar\hspace{3.45cm} 1--\pageref{LastPage}}}}
\fancyhead{}
\renewcommand{\headrulewidth}{0pt} 
\renewcommand{\footrulewidth}{0pt}
\setlength{\arrayrulewidth}{1pt}
\setlength{\columnsep}{6.5mm}
\setlength\bibsep{1pt}

\makeatletter 
\newlength{\figrulesep} 
\setlength{\figrulesep}{0.5\textfloatsep} 

\newcommand{\topfigrule}{\vspace*{-1pt}%
\noindent{\color{cream}\rule[-\figrulesep]{\columnwidth}{1.5pt}} }

\newcommand{\botfigrule}{\vspace*{-2pt}%
\noindent{\color{cream}\rule[\figrulesep]{\columnwidth}{1.5pt}} }

\newcommand{\dblfigrule}{\vspace*{-1pt}%
\noindent{\color{cream}\rule[-\figrulesep]{\textwidth}{1.5pt}} }

\makeatother

\twocolumn[
  \begin{@twocolumnfalse}
{\includegraphics[height=30pt]{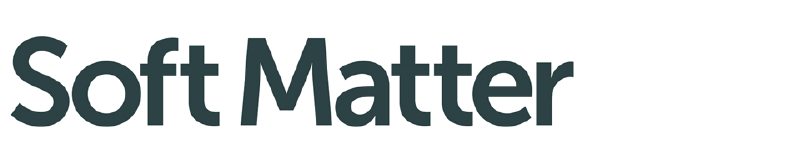}\hfill\raisebox{0pt}[0pt][0pt]{\includegraphics[height=55pt]{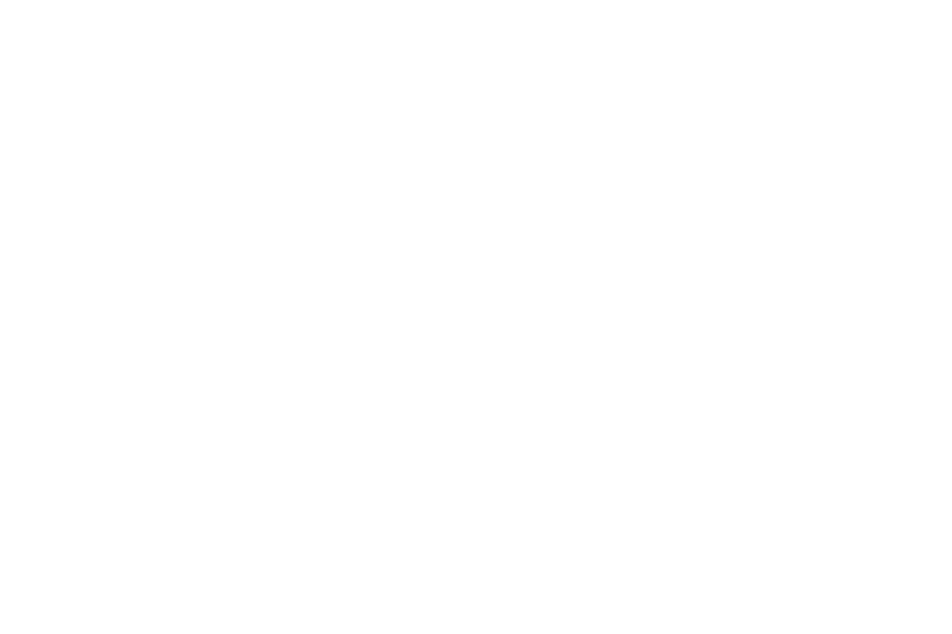}}\\[1ex]
\includegraphics[width=18.5cm]{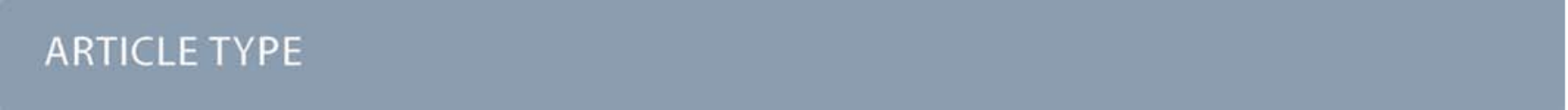}}\par
\vspace{1em}
\sffamily
\begin{tabular}{m{4.5cm} p{13.5cm} }

\includegraphics{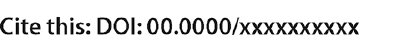} & \noindent\LARGE{\textbf{Fingering instability in spreading epithelial monolayers: roles of cell polarisation, substrate friction and contractile stresses}} \\
\vspace{0.3cm} & \vspace{0.3cm} \\

 & \noindent\large{Carolina Trenado,\textit{$^{a}$} Luis L. Bonilla,\textit{$^{a}$} and Alejandro Mart\'inez-Calvo\textit{$^{b,c,d}$}} \\

\includegraphics{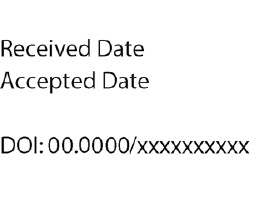} & \noindent\normalsize{Collective cell migration plays a crucial role in many developmental processes that underlie morphogenesis, wound healing, or cancer progression. In such coordinated behaviours, cells are organised in coherent structures and actively migrate to serve different biological purposes. In some contexts, namely during epithelial wound healing, it is well known that a migrating free-edge monolayer develops finger-like instabilities, yet the onset is still under debate. Here, by means of theory and numerical simulations, we shed light on the main mechanisms driving the instability process, analysing the linear and nonlinear dynamics of a continuum compressible polar fluid. In particular, we assess the role of cell polarisation, substrate friction, and contractile stresses. Linear theory shows that it is crucial to analyse the perturbation transient dynamics, since we unravel a plethora of crossovers between different exponential growth rates during the linear regime. Numerical simulations suggest that cell-substrate friction could be the mechanism responsible for the formation of complex finger-like structures at the edge, since it triggers secondary fingering instabilities and tip-splitting phenomena. Finally, we obtain a critical contractile stress that depends on cell-substrate friction and the initial-to-nematic length ratio, characterising an active wetting-dewetting transition. In the dewetting scenario, the monolayer retracts and becomes stable without developing finger-like structures.} \\


\end{tabular}

 \end{@twocolumnfalse} \vspace{0.6cm} 
  ]

\renewcommand*\rmdefault{bch}\normalfont\upshape
\rmfamily
\section*{}
\vspace{-1cm}


\footnotetext{\textit{$^{a}$~Department of Mathematics, Gregorio Mill\'an Institute, Fluid Dynamics, Nanoscience and Industrial Mathematics, Universidad Carlos III de Madrid, 28911 Legan\'{e}s, Spain; E-mail: ctrenado@ing.uc3m.es}}
\footnotetext{\textit{$^{b}$~Grupo de Mec\'anica de Fluidos, Gregorio Mill\'an Institute, Fluid Dynamics, Nanoscience and Industrial Mathematics, Universidad Carlos III de Madrid, 28911 Legan\'{e}s, Spain; E-mail: amcalvo@ing.uc3m.es}}
\footnotetext{\textit{$^{c}$~Princeton Center for Theoretical Science,  Princeton  University,  Princeton,  NJ  08544, USA}}
\footnotetext{\textit{$^{d}$~Department  of  Chemical  and  Biological  Engineering,  Princeton  University,  Princeton,  NJ  08544, USA}}





\section{Introduction}\label{sec:intro}
\begin{figure*}[htp]
    \centering
    \includegraphics[width=0.95\textwidth]{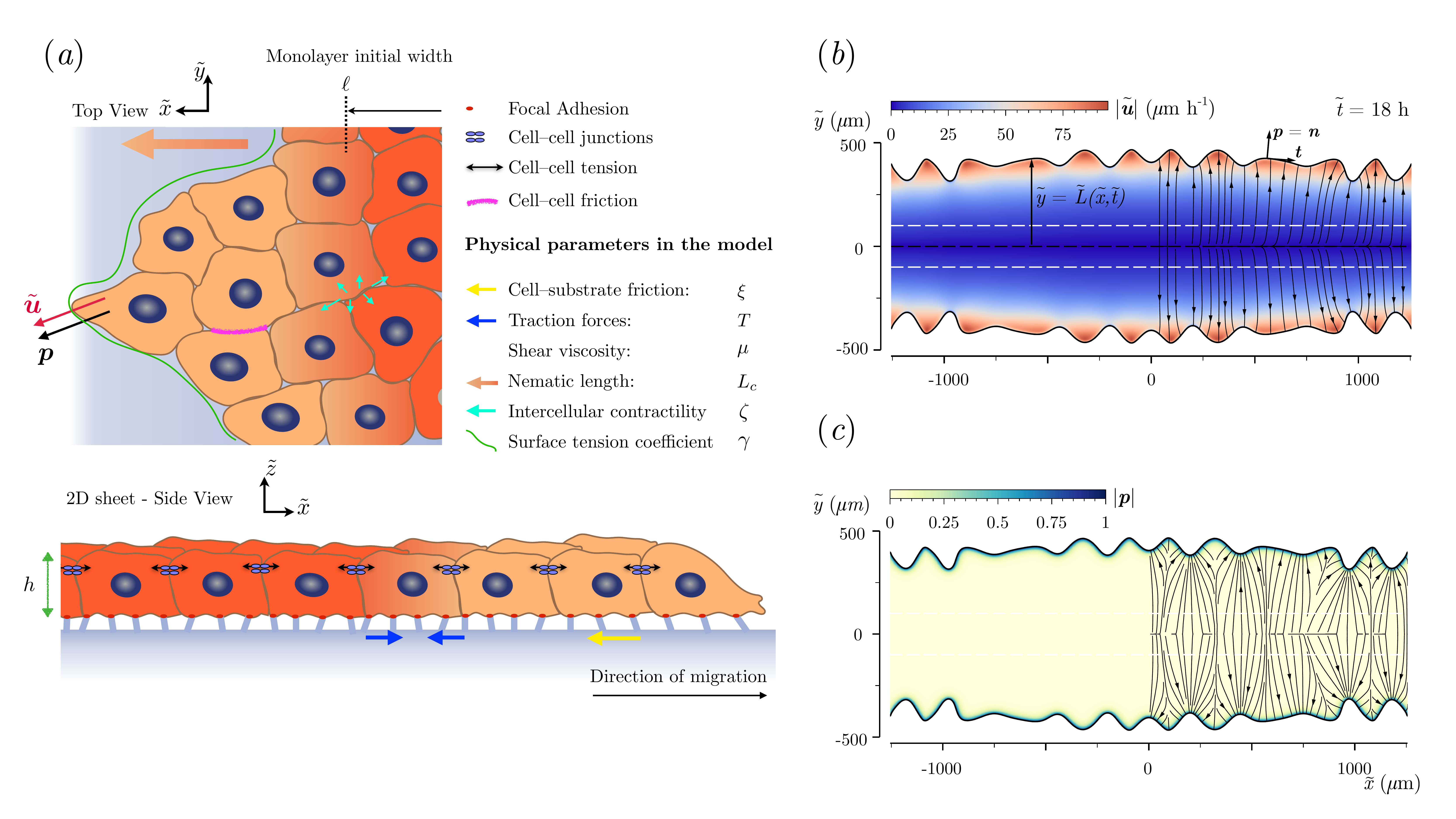}
    \caption{(a) Sketch showing the side and top views of a migrating epithelial monolayer. The different mechanisms and forces considered in the continuum model are indicated. (b,c) Snapshots showing the fingering instability in a spreading tissue monolayer obtained from the time-dependent numerical simulations of the complete equations of motion~\eqref{eq:pol}-\eqref{eq:initial} for $\xi = 1.5$ Pa\,min\,$\mu$m$^{-2}$, $\zeta = -2.8 $ kPa, $h = 5$ $\mu$m, $T = 1$ kPa, $\mu = 10^4 $ Pa min, $\ell = 100$ $\mu$m, and $L_c =12.5 $ $\mu$m ($\Lambda = 8$, $\beta = 1.5$, and $\alpha = -1.4$) at $\tilde{t} = 18$ h. The colourplots display the velocity and polarisation moduli $|\tilde{\boldm{u}}|$, $|\boldm{p}|$.}
    \label{fig:fig0}
\end{figure*}
Viscous fingering instabilities have been extensively studied since the seminal works of S. Hill~\citep{hill1952channeling}, R.L. Chuoke and coworkers~\cite{chuoke1959instability}, P.G. Saffman and G.I. Taylor~\citep{saffman1958penetration}, who observed and deduced that the interface between a moving fluid displacing another more viscous fluid in a Hele-Shaw cell~\citep{hele1898} is unstable to small disturbances. The reader is referred to refs.~\cite{bensimon1986viscous,homsy1987viscous,kessler1988pattern,oron1997long,ben2000cooperative} for detailed excellent reviews on hydrodynamic finger-like instabilities and pattern formation. Such interfacial instabilities have been also observed in several biological contexts during \textit{in vivo} processes. In particular, fingering instabilities arise both, in prokaryotic systems, namely during bacterial growth~\citep{ben1992adaptive,ben1994generic,shapiro1995significances,kearns2010field,ben2012bacterial,farrell2013mechanically}, and in eukaryotic group of cells, for instance during collective cell migration, which drives a myriad of crucial biological processes such as morphogenesis~\cite{lecuit2011force,goodwin2019smooth,nerger2021local}, embryogenesis~\cite{forgacs_newman_2005,lecuit2007cell}, or tumour invasion~\cite{abercrombie1962surface,friedl2009collective,streitberger2020tissue,moitrier2019collective,bonilla2020tracking}. Regarding collective cell migration, a huge interdisciplinary effort has been devoted to unravel the main physical mechanisms involved in these complex group behaviours, i.e. how cell-cell interactions can give rise to a cohesive coordinated motion~\citep{mayor2016front,stramer2017mechanisms,ladoux2017mechanobiology,xi2019material,alert2020physical}. Nonetheless, the key mechanisms driving the emergence of fingering instabilities during collective cell migration is still under debate. As we detail below, the onset of instability might be a consequence of the complex coordinated behaviour of cells, involving the interplay between mechanical and chemical signals, and the interaction with their environment. 

'Cell' was the word coined by the big-name British polymath Robert Hooke to name these closed units that form a complex organised structure~\citep{hooke2003micrographia}. In particular, the interaction between them and with the medium they inhabit, gives rise to highly correlated collective motion and mechanical stresses~\cite{ladoux2017mechanobiology}. As a group, they usually become motile under chemical and mechanical stimuli, enabling development and regeneration processes. For instance, in embyrogenesis and morphogenesis, cells grow significantly until they conform a global morphology, during wound healing, cells migrate to close a gap, and also in pathological processes such as tumor invasion, malignant cells migrate and grow invading healthy populations of cells. 

In 1962, M. Abercrombie and E. Ambrose wrote 'It is a well-known principle that epithelium will not tolerate a free edge'~\citep{abercrombie1962surface}, meaning that, a free-edge two-dimensional epithelial sheet resting on a substratum, will migrate by an active pulling action of polarised cells located near the edge, until they reach a boundary or another epithelium. During this process, wound healing \textit{in vivo} and \textit{in vitro} experiments have revealed that the epithelium edge can form multi-cellular finger-like structures as it migrates to close the wound. Hence, wound healing is a very convenient biological process to study collective cell migration and the formation of fingering patterns. Nonetheless, the physical mechanisms responsible for such instability remain controversial. Some authors, by means of experiments~\citep{omelchenko2003,poujade2007,mayor2016front,begnaud2016mechanics,vishwakarma2018mechanical,xi2019material,yang2020leader}, theory and numerical simulations of continuum~\citep{williamson2018stability,vishwakarma2018mechanical,yang2020leader} and agent-based models~\citep{sepulveda2013,tarle2015modeling,yang2020leader}, argue that the existence of polarised leader cells at the edge and the signaling with their followers, may drive the emergence of finger-like protrusions by using an active pulling action. In a recent work~\citep{bonilla2020tracking}, numerical simulations of the active-vertex model suggest that leader cells are not necessary to trigger the fingering instability if the dynamics of cell centers includes collective tissue forces, velocity alignment, and inertia, which are the key mechanisms driving the instability. In a few recent works, linear stability analyses of continuum active fluid models have been derived to explain the experimental observations, arguing that the emergence of fingering instabilities could be explained by means of a kinematic hydrodynamic-like interfacial instability~\citep{basan2013,nesbitt2017edge,Alert2019,yang2020leader}. 



Concerning the latter approach, this class of theoretical models aim to describe the epithelial monolayer as an active fluid or visco-elastic continuum by means of macroscopic fields such as velocity, displacement, cell density, and cell polarisation. Refs.~\cite{zimmermann2014instability,nesbitt2017edge,bogdan2018fingering} study the linear stability of a two-dimensional incompressible active fluid modelled via Toner-Tu-like equations, in strip~\citep{zimmermann2014instability,nesbitt2017edge} and circular~\citep{bogdan2018fingering} configurations. The former works show that a quiescent monolayer is unstable to small perturbations for some ranges of wavenumbers depending on inertial effects, whereas a monolayer moving with uniform velocity is stable to all wavenumbers. Regarding the circular configuration, ref.~\citep{bogdan2018fingering} shows that an initially migrating monolayer is unstable when cell growth is taken into account. More recently, the work of~\cite{yang2020leader} shows that the uniform-velocity migrating monolayer is unstable when considering leader cells at the interface, in the absence of cell growth. In this work, leader cells are taken into account within the continuum framework by introducing a curvature-dependent force at the interface, similarly to the work of~\cite{mark2010physical}, where the interface is modelled as an elastic membrane with bending resistance via the Helfrich-Canham potential~\citep{Canham1970,helfrich1973elastic,zhong1987instability,Zhong1989,Seifert1993,seifert1997configurations}.

In a series of recent works~\citep{blanch2017effective,alert2018role,moitrier2019collective,Alert2019,perez2019active,alert2020physical,heinrich2020size}, the epithelial monolayer is modelled as a compressible viscous polar fluid, taking into account cell-substrate active traction forces, cell-substrate friction, viscous, surface tension, and contractile forces. Ref.~\cite{Alert2019} shows that the flat-front solution becomes unstable by means of a kinematic mechanism: when a small perturbation is introduced at the edge, the gradient of velocity across the monolayer, which comes from the balance between contact-active and viscous forces, makes the crests to move faster than the valleys. However, these works do not explore the time-dependent stability of the migrating monolayer, something that we believe is crucial to assess the stability and the most amplified wavelengths, since the velocity profile is not uniform and the velocity at the edge is not generically constant. Additionally, the effect of these forces in the nonlinear regime remains almost unexplored, which is essential to unravel the main mechanisms driving the instability and to assess if this kind of active continuum modelling is able to qualitatively reproduce the experimental observations. Other works have explored the long-time behaviour of spreading monolayers via numerical simulations of similar continuum models~\citep{lee2011crawling,kopf2013continuum}. Ref.~\cite{lee2011crawling} takes into account time-dependent, inertial and corotational effects in cell polarisation, as well as viscoelastic effects via the Maxwell model. Ref.~\cite{kopf2013continuum} considers a neo-Hookean elastic material taking into account cell proliferation, stress-polarisation coupling, as well as long-range chemo-mechanical interactions, i.e. a feedback loop between tissue deformation and inter-cellular chemical signaling. Both works obtain finger-like patterns at the edge, some of them qualitatively similar to the ones observed in the experiments. Nonetheless, these works do not systematically explore the role of the different forces involved during the migration, and which are the key mechanisms underlying these complex patterns in the nonlinear regime. 

Hence, by using the same continuum framework as in refs.~\citep{blanch2017effective,alert2018role,moitrier2019collective,Alert2019,perez2019active,alert2020physical,heinrich2020size}, our work aims to distill the role of viscous, active tractions, contractile forces and cell-substrate friction, which are ubiquitous during the migration of epithelial monolayers, on the linear and nonlinear regimes, by analysing the complete time-dependent linear stability, and via time-dependent numerical simulations to explore the long-time behaviour.

The paper is organised as follows. In Sec~\ref{sec:sec2} we describe the active continuum model to analyse the collective dynamics of a two-dimensional cell sheet. In Sec~\ref{sec:sec3}, we briefly describe the procedure to obtain the base-flow and the corresponding linear stability analysis of such flow configuration. In Sec~\ref{sec:sec4}, we compare the results obtained from the linear stability analysis and the time-dependent two-dimensional numerical simulations of the complete equations of motion. Using both frameworks, we compare the numerical simulations of the spreading epithelial monolayer with experiments reported in the literature, and we also consider different distinguished limits to discuss the role of the governing physical parameters on the migration, stability, and spreading nonlinear regimes of the epithelial sheet. Conclusions are drawn in Sec~\ref{sec:conclusions}, where we also discuss future theoretical avenues in understanding the collective behaviour of epithelial monolayers. Appendix~\ref{app:appendix} contains a brief derivation of the continuum model used in the present work.

\section{Theoretical modeling}\label{sec:sec2}
\begin{table*}[h!]
\begin{tabular}{p{5.2cm}p{7.5cm}p{3cm}p{1cm}}
 \hline 
 \hline
\hspace{0.1cm} Physical & \hspace{0.7cm}Definition & Range of values\\
Parameters &  & \\
 \hline
 $\ell$    & Monolayer initial width    & $100$--$400$ $\mu$m & ~\citep{blanch2017effective,Alert2019}\\
 $\mu$     & Shear viscosity  & $10^4$--$10^6$ Pa min & ~\citep{blanch2017effective,perez2019active}\\
 $h$       & Tissue height & 5 $\mu$m & ~\citep{trepat2009physical,perez2019active} \\
 $T$       & Traction coefficient & 0.1--0.5 kPa &~\citep{blanch2017effective,perez2019active}\\
 $L_c$     & Nematic length & 20--100 $\mu$m & ~\citep{blanch2017effective,perez2019active}\\
 $\zeta$   & Intercellular contractility & -$5$-- -$20$ kPa & ~\citep{perez2019active}\\
 $\xi$     & Cell-substrate friction & $0.2$--$4$ Pa\,min\,$\mu$m$^{-2}$ & ~\citep{blanch2017effective}\\
 $\gamma$ & Surface tension coefficient &  $1$--$10$ mN m$^{-1}$ &~\citep{foty1994liquid} \\
 \hline
 \hline
 Dimensionless    &  & \\
\,\,\, Parameters    &  & \\\hline
 $\Lambda = \ell/L_{\textrm{c}}$ & Initial-to-nematic length ratio & 1 -- 8\\
 $\beta = \xi \ell^2/\mu$ & Friction-to-viscous forces ratio  & 0.04 -- 3.2\\
 $\alpha = \zeta h/(\ell T)$  & Contractile-to-active-contact forces ratio & -10 -- -0.25\\
$\Ca~=~\gamma h/(\ell^2 T)$ & Active Capillary number & 10$^{-4}$ -- 0.05 \\ 
\hline
\end{tabular}
\ \\
\caption{Estimates and experimental measurements of the physical parameters and estimated values of the dimensionless numbers.\label{tab:table1}}
\end{table*}

To analyse the collective cell dynamics of a spreading epithelial monolayer we adopt the same theoretical continuum framework as in refs.~\citep{blanch2017effective,alert2018role,Alert2019,perez2019active,moitrier2019collective,alert2020physical,heinrich2020size}, mainly inspired by active-gel physics~\citep{prost2015active} and liquid crystal theory~\citep{de1993}. In particular, we consider an active polar fluid surrounded by a passive ambient, propagating in the direction perpendicular to its interface due to cell-substrate active traction forces (see the sketch in Fig.~\ref{fig:fig0}a). These forces are exerted by cells' actomyosin cytoskeleton at focal adhesion sites generating contractile forces that are transmitted to the group enabling the collective migration. The flow is described in terms of the polarity field $\boldm{p}$, which takes into account the polarity alignment and direction of cells, and the velocity field $\boldm{u}$. We assume that the flow is two-dimensional, thus $\boldm{u}$ and $\boldm{p}$ are depth-average fields. This assumption provides a good approximation if the slip length is significantly larger than the layer thickness~\citep{deGennes1979,Brochard1992,Brochard1997dewetting}, which may be the case for thin epithelial monolayers. We further assume that the flow is compressible, and neglect inertial effects and internal pressure due to cell proliferation. The spontaneous spreading process is triggered by polar cell-substrate active traction forces, which are assumed to be proportional to $\boldm{p}$, and in balance with viscous forces. Within the stress balance, we also consider cell-substrate friction and contractile forces, disregarding flow alignment, and the elastic response of the tissue. For the sake of simplicity, the shear and bulk viscosity coefficients are assumed to be equal. For cell polarisation dynamics, we consider a diffusion-dominated polarisation field, neglecting advection, corotation, and flow alignment effects, i.e. negligible back-coupling of the flow on the polarisation field.

The above physical mechanisms aim to explain the complex behaviours arising in a migrating epithelial monolayer, namely the formation of finger-like structures at the edge during the spreading process. 


To non-dimensionalise the equations of motion, we assume that viscous and active traction forces are in balance, which yields, $\mu \partial_x u_y \sim T/h$, where $u_y$ is the velocity across the monolayer (see Fig.~\ref{fig:fig0}a), $\mu$ denotes the shear viscosity coefficient, $h$ is the height of the tissue measured from the
substrate, and $T$ is a traction coefficient accounting for cell-substrate active forces of polarised cells. Hence, the
characteristic velocity reads, $u_c \sim T \ell^2/(\mu h)$, where $\ell = \bar{L}(t = 0)$ is the initial width of the monolayer. It proves to be convenient to introduce the following non-dimensionalization
\begin{equation}\label{eq:scaling}
\boldm{x} = \frac{\bar{\boldm{x}}}{\ell}, \quad \boldm{u} = \bar{\boldm{u}} \frac{\mu h}{\ell^2 T}, \quad t = \bar{t} \frac{\ell T}{\mu h}, \quad L = \frac{\bar{L}}{\ell}
\end{equation}
where bars denote dimensional variables, $\boldm{x}$, $t$, and $L$, are the position vector, time, and the monolayer half-width, respectively.

At this point it is important to emphasize that the continuum model does not describe explicitly subcellular length scales, size changes, and shape variations of the cells~\citep{alert2020physical}. These lengths and effects are considered implicitly in the model through the polarity field $\boldm{p}(\boldm{x},t)$, which is independent of the flow and satisfies,
\begin{equation}\label{eq:pol}
\bnabla^2 \boldm{p} = \Lambda^2 \boldm{p},
\end{equation}
where $\Lambda = \ell/L_c$ is the dimensionless parameter comparing the initial width of the monolayer $\ell$, and the characteristic decay length of the polarity field, denoted by $L_c$. 


The dynamics of the epithelial monolayer is described through the following inertialess stress balance,
\begin{equation}\label{eq:mom}
\bnabla \bcdot \boldm{\sigma} + \boldm{f} = \boldm{0}.
\end{equation}
The cell polarity field modifies the stress tensor $\boldm{\sigma}$ (monolayer tension) and the external body forces $-\boldm{f}$ (cell-substrate traction stresses), producing active contractile stresses and active contact forces, respectively. The stress tensor contains viscous and contractile terms, $\boldm{\sigma} = \bnabla \boldm{u} + (\bnabla \boldm{u})^{\text{T}} - \alpha \, \boldm{p} \boldm{p}$, where $\alpha < 0$ is the dimensionless signature of the active stress. The external body force reads: $\boldm{f} = -\beta \boldm{u} + \boldm{p}$, which takes into account cell-substrate friction and active traction forces, respectively. In particular, the coarse-grained friction and active tractions are assumed to be spatially uniform for simplicity. The dimensionless parameters are: $\alpha= \zeta h/(\ell T)$ and $\beta = \xi \ell^2/\mu$, which compare inter-cellular contractile forces with active traction forces, and friction forces with viscous forces, respectively, where $\zeta$, and $\xi$ are the dimensional contractility and cell-substrate friction coefficients. Furthermore, the dimensionless number $\beta$ is also defined as the squared ratio between the initial length of the monolayer, $\ell$, and the screening length, $\sqrt{\mu/\xi}$, which characterises the penetration of the viscous stress inside the monolayer. 

We impose the following kinematic boundary condition:
\begin{equation}\label{eq:kinematic}
(\partial_t \boldm{x}_s - \boldm{u}) \bcdot \boldm{n} = 0 \quad \text{at} \quad y = L(x,t),
\end{equation}
where $\boldm{x}_s$ is the parameterisation of the interface, and $\boldm{n}$ its unit normal vector. Additionally, we also impose that the polarisation field is normal to the interface, and the surface stress balance:
\begin{equation}\label{eq:stress}
\boldm{p} = \boldm{n}, \quad \text{and} \quad \boldm{\sigma} \bcdot \boldm{n} + \Ca (\bnabla_s \bcdot \boldm{n})\boldm{n} = \boldm{0} \quad \text{at} \quad y = L(x,t),
\end{equation}
where $\bnabla_s \bcdot \boldm{n} = -\partial_x^2 L [1+(\partial_x L )^2]^{-3/2}$ is twice the mean curvature of the interface, $\bnabla_s = (\mathbfsf{I} - \boldm{n}\boldm{n})\bcdot \bnabla$ is the surface gradient operator, and $\Ca = \gamma h/(\ell^2 T)$ is an active Capillary number based on active traction forces. For the sake of simplicity, we have neglected the effect of interfacial tension, although as shown in Table~\ref{tab:table1}, it may become relevant in some situations. Equivalently to the classical hydrodynamic instability, surface tension has a stabilising effect, as shown in ref.~\citep{Alert2019}, by generating an interfacial stress that tries to flatten the initial perturbation at the edge. Hence, the governing dimensionless parameters are $\Lambda$, $\beta$, and $\alpha$, whose typical values are given in Table \ref{tab:table1}, using different estimates and experimental measurements of the physical parameters available in the literature.




\begin{figure}[ht!]
    \centering
    \includegraphics[width=0.48\textwidth]{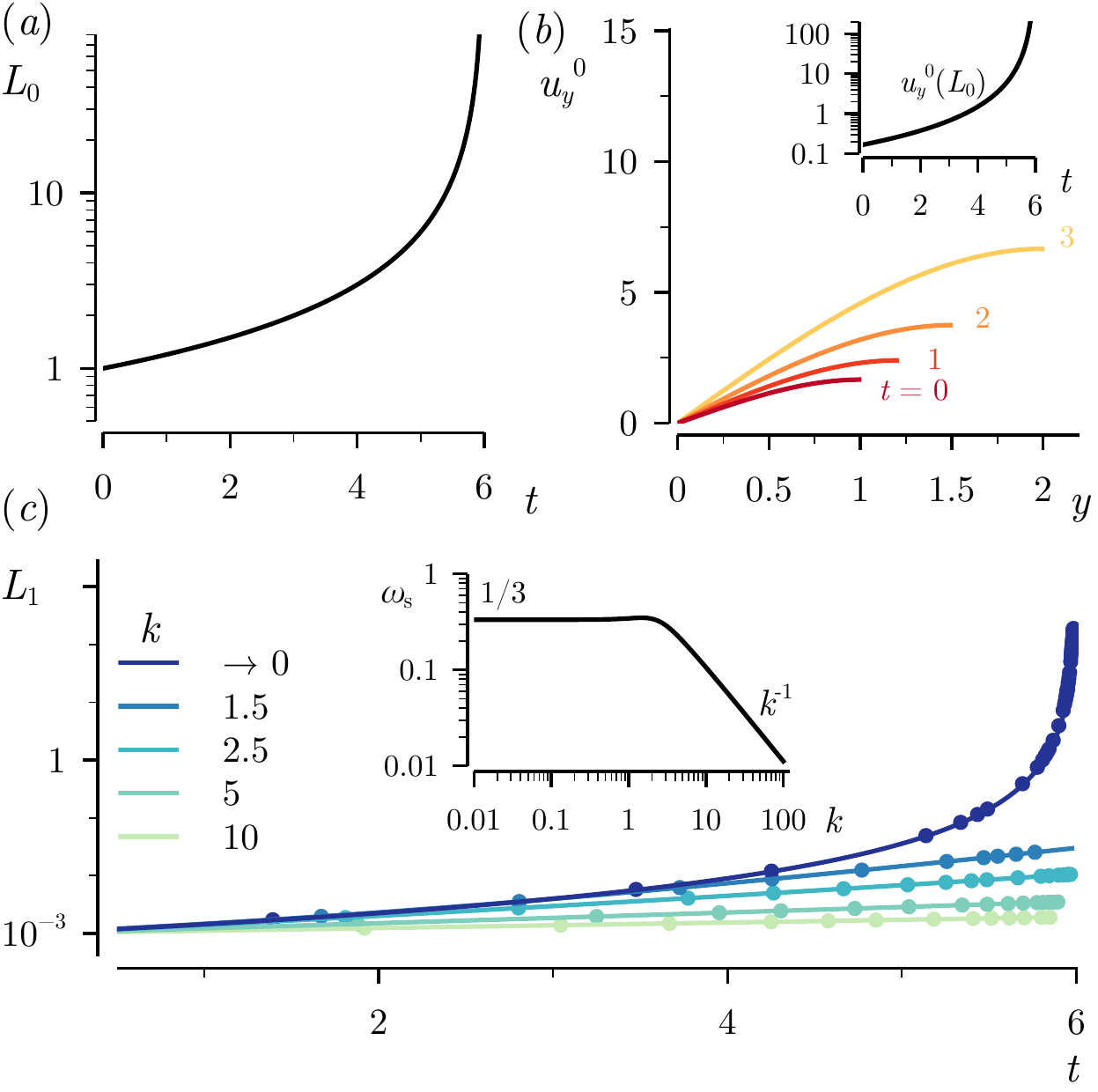}
    \caption{Limit $\Lambda=\alpha=\beta=0$. (a) Position of the base-flow interface $L_0$ as a function of time $t$. (b) Base-flow velocity $u_y^0$ as a function of $y$ for different times. The inset shows the velocity at the edge of the monolayer $y = L_0$ as a function of time $t$. (c) Perturbation amplitude $L_1$ as a function of time $t$, for different values of the wavenumber $k$ indicated in the legend. The inset shows the short-time exponential growth rate $\omega_{\textrm{S}}$ as a function of $k$~\citep{Alert2019}.}
    \label{fig:fig1}
\end{figure}

We numerically integrate eqns~\eqref{eq:pol}-\eqref{eq:stress} imposing that the monolayer is not polarised at the centre line, together with symmetry boundary conditions:
\begin{equation}
p_y = \partial_y p_x = u_y = \partial_y u_x = 0 \quad \text{at} \quad y = 0, \label{eq:bc_symmetry}
\end{equation}
and symmetric boundary conditions at the planes $x = 0$ and $x = \pi/k$. To trigger the fingering instability, the interface of the spreading monolayer is slightly perturbed by a harmonic disturbance at $t = 0$,
\begin{equation}\label{eq:initial}
L(x,0) = 1 - L_{1,0} \cos(k x),    
\end{equation}
for $0 \leq x \leq \pi/k$, where $k$ is the wavenumber, and $L_{1,0} \ll 1$ is the initial amplitude of the perturbation. The initial conditions for $\boldm{p}$ and $\boldm{u}$ are not imposed, since we employ a standard Newton-Raphson root-finding algorithm to obtain them with the above boundary conditions. For the time discretisation, we use a variable-step BDF method with 2/5 variable order, and the Arbitrary Lagrangian-Eulerian method within a finite-element framework to numerically solve the time-dependent free-boundary problem. In particular, the numerical technique is the same as the one employed in refs.~\cite{martinez2020natural,moreno2020stokes}, where the reader can find further technical details.

\section{Linear Stability Analysis}\label{sec:sec3}

\begin{figure*}[h!]
    \centering
    \includegraphics[width=0.95\textwidth]{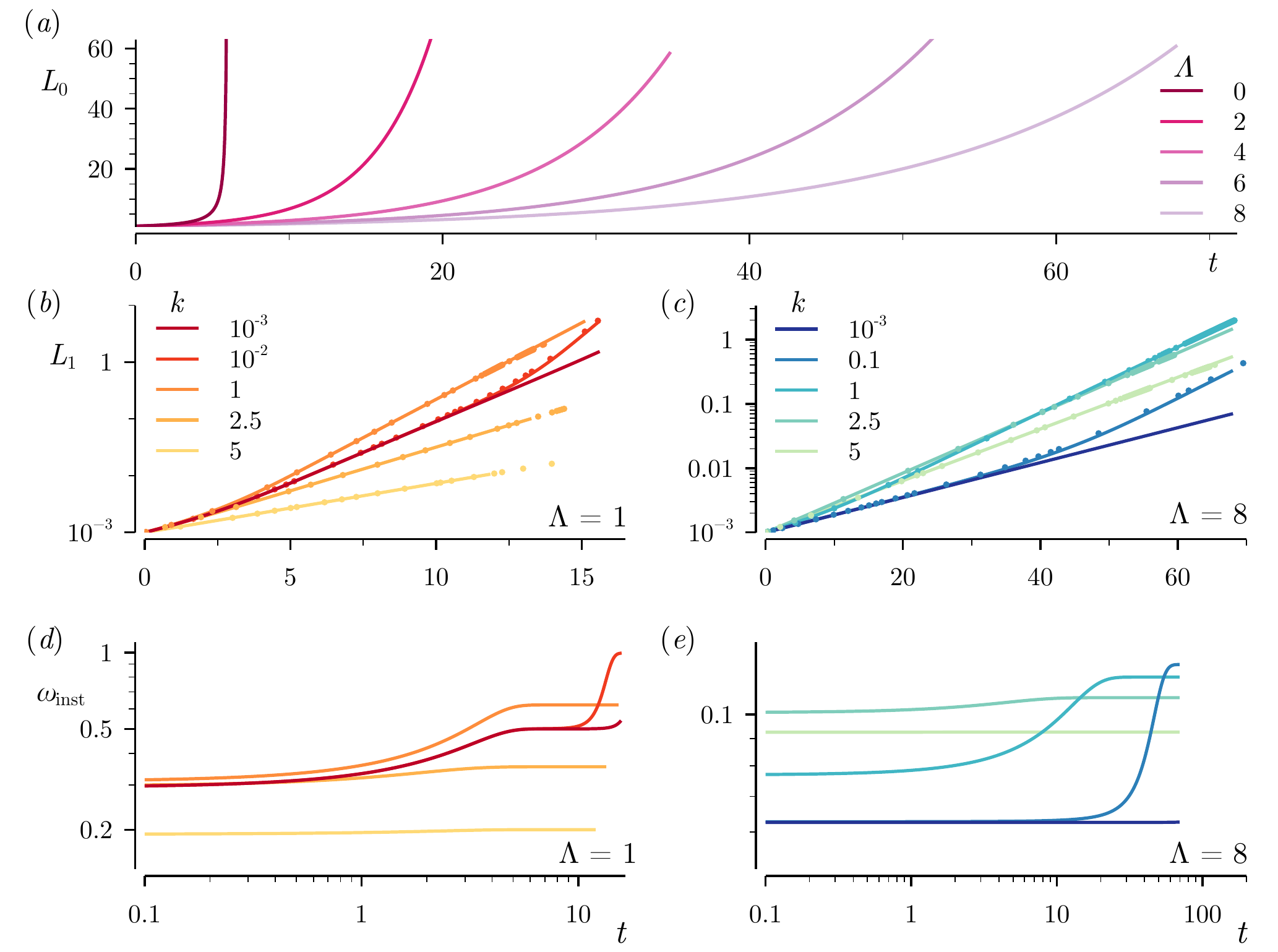}
    \caption{(a) Base-flow interface position $L_0$ corresponding with the flat-front solution as a function of time $t$, for different values of $\Lambda$ indicated in the legend. (b,c) Perturbation amplitude $L_1$ as a function of time $t$ for (a) $\Lambda = 1$ and (b) $\Lambda = 8$, and different values of the wavenumber $k$ indicated in the legends. (c,d) Instantaneous exponent $\omega_{\textrm{inst}} = \textrm{d} \ln(L_1)/\textrm{d}t$ as a function of time $t$ for the same values of $\Lambda$ and $k$ as in panels (a,b).}
    \label{fig:fig2}
\end{figure*}

To analyse the migration and the stability of the epithelial monolayer, we perform a linear stability analysis considering the configuration shown in Fig.~\ref{fig:fig0}(a), which is commonly used in wound healing assays. To this end, we linearise the equations of motion around a certain base state and analyse the growth of small perturbations. In particular, we consider the unidirectional flat-front solution as base flow~\citep{Alert2019}, which reads: $\boldm{p}_{0} = p_y^0(y) \boldm{e}_y$ for the polarisation field, and $\boldm{u}_0 = u_y^0(y) \boldm{e}_y$, for the velocity field. The detailed expressions of the base flow depending on the dimensionless parameters are given below. Once obtained $\boldm{p}_0$ and $\boldm{u}_0$, the position of the base-flow interface, $L_0(t)$, is determined according to the kinematic condition~\eqref{eq:kinematic}, which simplifies to:
\begin{equation}\label{eq:baseflow_kinematic}
\frac{\textrm{d}L_0(t)}{\textrm{d} t} =u_y^0(y=L_0(t)).
\end{equation}
To address the stability of the flat-front spreading monolayer, all the variables are perturbed around such base flow by small-amplitude disturbances. It is important to emphasise that the base state varies with time, thus the coefficients of the linearised equations of motion are also functions of time, which precludes the assumption of a simple exponential growth of disturbances. Therefore, the time-dependent perturbations are decomposed as Fourier wave-like modes defined by the wavenumber $k$:
\begin{align}\label{eq:normalmode}
(\boldm{p}, \boldm{u},L) = (\boldm{p}_0,\boldm{u}_0,L_0)+(\boldm{p}_1, \boldm{u}_1,L_1) \exp(i k x),
\end{align}
where $|\boldm{p}_1|$, $|\boldm{u}_1|$, $|L_1| \ll 1$. This means that an algebraic dispersion relation, $D(k,\omega)~=~ 0$, between an exponential growth rate of perturbations $\omega$ and the wavenumber $k$, cannot be derived. 


Introducing eqn~\eqref{eq:normalmode} into the system of eqns~\eqref{eq:pol}--\eqref{eq:mom}, allows us to solve $\boldm{p}_1$ and $\boldm{u}_1$, with the appropriate linearised boundary conditions~\eqref{eq:stress} and~\eqref{eq:bc_symmetry}. Finally, the evolution of the interface perturbation $L_1$ is obtained from the linearised kinematic condition~\eqref{eq:kinematic}:
\begin{equation}\label{eq:eqL1}
\frac{\text{d} L_1}{\text{d} t} = \partial_y u_y^0 \, L_1+u_y^1 =  \omega(k,t) L_1 \quad \text{at} \quad y = L_0(t),
\end{equation}
where we have taken advantage that $u_y^1$ is proportional to $L_1$, to define $\omega(k,t)$ as the time-dependent growth rate. This allows us to solve eqn~\eqref{eq:eqL1}, which yields~\citep{Alert2019}
\begin{equation}
L_1(t) = L_{1,0} \exp \left( \int_0^t \omega(k,t) \textrm{d} t\right). \label{eq:L1_sol}
\end{equation}
Hence, to study the linear stability of the spreading front, we shall analyse the transient dynamics of the perturbations.

\begin{figure*}[h!]
    \centering
    \includegraphics[width=0.95\textwidth]{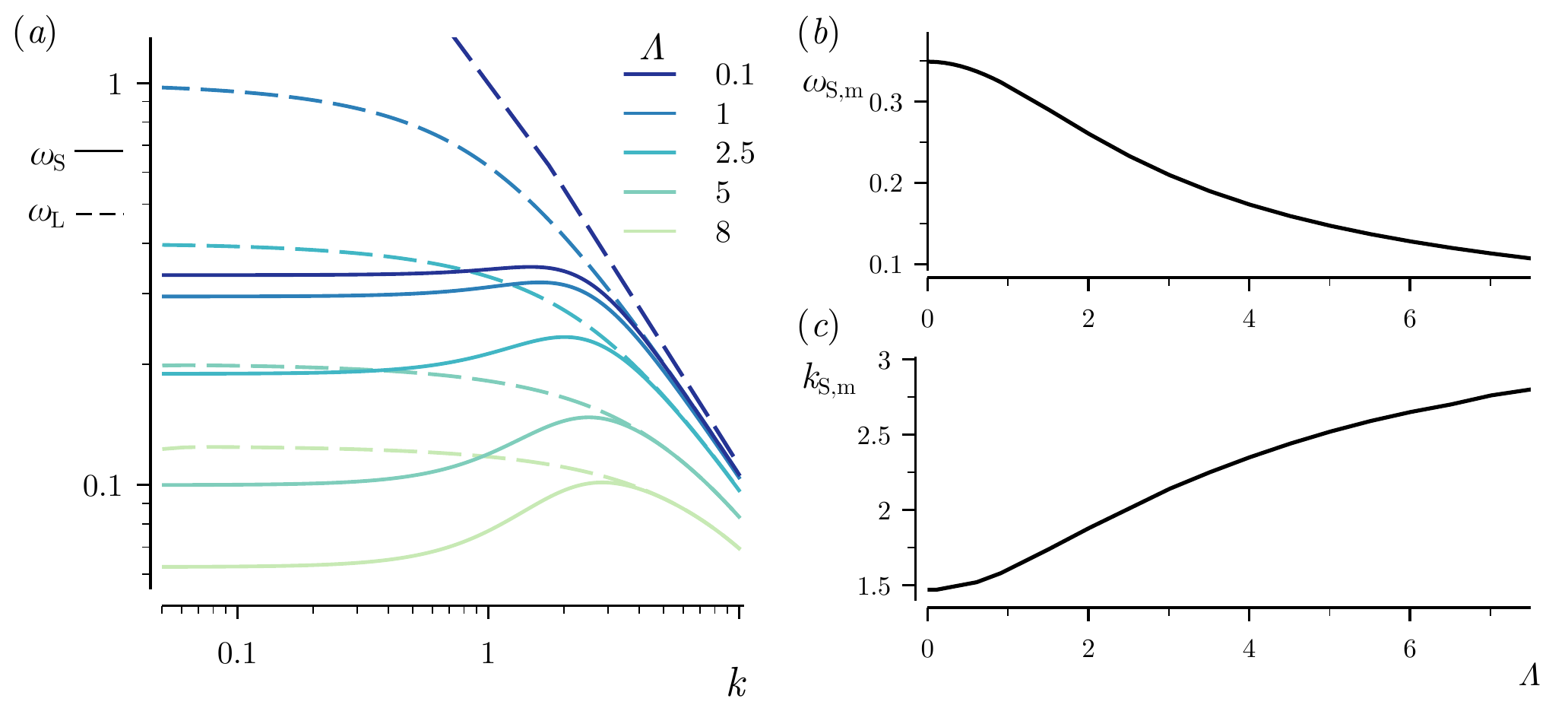}
    \caption{(a) Short-time $\omega_{\textrm{S}}$ and large-time exponential growth rate $\omega_{\textrm{L}}$ as functions of the wavenumber $k$ in solid and dashed lines, respectively. (b) Maximum short-time growth rate $\omega_{\textrm{S},\textrm{m}}$ and (c) the corresponding most amplified wavenumber $k_{\textrm{S},\textrm{m}}$ as functions of $\Lambda$.}
    \label{fig:fig3}
\end{figure*}

\section{Results and discussion}\label{sec:sec4}
In this section, we analyse the linear and nonlinear dynamics of the migrating epithelial monolayer. To this end, we employ the linear stability analysis described above and time-dependent numerical simulations of the complete set of eqns~\eqref{eq:pol}--\eqref{eq:initial}.

In particular, Fig.~\ref{fig:fig0}(b) illustrates the spreading process of an epithelial monolayer and the associated fingering instability by showing a snapshot at $\tilde{t} = 18$h obtained from a time-dependent numerical simulation of the system~\eqref{eq:pol}--\eqref{eq:initial}, for a combination of the physical parameters indicated in the caption of Fig.~\ref{fig:fig0} (with the following values of the dimensionless parameters: $\Lambda = 8$, $\beta = 1.5$, and $\alpha = -1.4$). This result evidences that this simplified continuum model is able to reproduce the finger-like structures observed in the experiments~\citep{poujade2007,petitjean2010velocity}.

Additionally, the velocity patterns of migrating epithelial monolayers has been deeply studied in \textit{in vitro} processes~\citep{poujade2007,petitjean2010velocity}. Experimental measurements show a significant difference in the velocity field between cells located at the edge of the monolayer, with average velocities close to 25$\pm$5 $\mu$m h$^{-1}$, and those in the bulk, with 5 $\mu$m h$^{-1}$. This phenomenon is also observed in our numerical simulations, as shown in Fig.~\ref{fig:fig0}(b), where the modulus of $\tilde{\boldm{u}}$ is significantly smaller in the central region of the monolayer than at the free edge. Additionally, several works argued that these finger-like patterns arise due to the dynamics of larger cells located at the edge of the monolayer, usually referred to as leader cells, which lose their epithelial characteristics and drag their neighbours to form fingers~\citep{poujade2007}. The continuum model considered in the present work, cannot account for these details and behaviors at the single-cell level. Nonetheless, these effects are incorporated at the supracellular scale through the polarisation field and the active traction forces. As seen in Fig.~\ref{fig:fig0}(c), cells located at the free edge are strongly polarised compared to those in the bulk, and thus exert a larger traction because the edge of the monolayer is not contact-inhibited. These forces are in turn transmitted through viscous forces to the bulk of the monolayer, which enables the collective migration. 

Moreover, finger-like instabilities appear typically after two hours in \textit{in vivo} experiments. Fig.~\ref{fig:fig0}(b) shows that, after $18$h, the length of the fingers in the numerical simulations is about 100 $\mu$m, which compares fairly well with those measured in experiments~\citep{poujade2007}. In experiments, after a few hours, even though most of the cells at the border of the epithelium have a perpendicular direction to the edge of the strip, some fingers may eventually develop an orientation which is not perpendicular to the initial straight monolayer. As it can be observed in Fig.~\ref{fig:fig0}(c), this effect is also captured by the continuum approximation, which shows that cells located at the free edge are polarised perpendicularly to the interface, and fingers eventually deviate from their initial orientation forming complex patterns.

As cells migrate collectively, large-scale deformation patterns arise and also velocity fluctuation fields show large-scale swirl patterns~\citep{petitjean2010velocity,angelini2010cell}. Experimental works have quantified the characteristic size of these swirls, which ranges from $200$ $\mu$m at low densities, to $350$ $\mu$m at high densities. The continuum model used here does not capture these two properties observed in the experiments at the collective-cell level, as shown in Fig.~\ref{fig:fig0}. Incorporating cell proliferation, cell density, or the rheological properties of the epithelial monolayer into the continuum approximation may reproduce these phenomena.

Here, we assess the role of the different physical parameters on the onset of such complex edge patterns observed in Fig.~\ref{fig:fig0}, by using both the linear stability analysis and the previously described time-dependent numerical simulations. To proceed with the stability analysis, we first compute the polarisation and velocity fields corresponding with the flat-front solution,
\begin{equation}
\boldm{p}_0 = \frac{\text{sinh}(\Lambda y)}{ \text{sinh}(\Lambda L_0)} \boldm{e}_y,
\label{eq:p_flat_complete}
\end{equation}
\begin{align}
& \boldm{u}_0 = \left[ \frac{\alpha \csch
(\Lambda L_0)}{2\Lambda} \left( \frac{\sinh(\Lambda y)}{\alpha \Lambda \left(\frac{\beta}{2 \Lambda^2} -1  \right)} - \frac{\csch(\Lambda L_0) \sinh(2 \Lambda y)}{\frac{\beta}{2 \Lambda^2} -4} \right)  + \right. \nonumber & \\
& \left. \frac{\alpha \sech
(\sqrt{\beta/2} L_0)}{2\sqrt{\beta/2}} \left(1 - \frac{\coth (\Lambda L_0)}{\alpha \Lambda \left(\frac{\beta}{2 \Lambda^2} -1  \right)} + \frac{2(1+\coth(\Lambda L_0)^2)}{\frac{\beta}{2\Lambda^2} -4}\right) \times \right. \nonumber & \\
& \left. \sinh(\sqrt{\beta/2} y) \right] \boldm{e}_y.
\label{eq:u_flat_complete}
\end{align}
The evolution of $L_0(t)$ can be then obtained using the kinematic condition~\eqref{eq:baseflow_kinematic}. We now extract the role of the different physical forces involved in the spreading process by considering several distinguished limits. These limits are presented in order of increasing complexity.

\subsection{Parameter-free solution: balance of viscous and active traction forces}\label{subsec:case_1}

Here, we consider the distinguished limit where $\Lambda = \beta = \alpha = 0$. In this limiting case, the base-flow polarity and velocity fields simplify to 
\begin{equation}\label{eq:baseflow_lambda0}
\boldm{p}_0 = \frac{y}{L_0} \boldm{e}_y, \quad \text{and} \quad \boldm{u}_0 = \frac{y L_0}{4}\left(1- \frac{y^2}{3 L_0^2} \right) \boldm{e}_y.
\end{equation}
Hence, the evolution of $L_0(t)$ can be straightforwardly obtained via the kinematic condition~\eqref{eq:baseflow_kinematic}, $\textrm{d}L_0/\textrm{d}t = L_0^2/6$, which yields: $L_0(t) = (1-t/6)^{-1}$. The temporal evolution of $L_0$ and $\boldm{u}_0$ is shown in Figs.~\ref{fig:fig1}(a,b). This simplified base-flow solution grows linearly with time for $t \ll 1$, but experiences a singularity at $t = 6$ where $L_0$ and the edge velocity $u_y^0(L_0)$ diverge. This unphysical behaviour simply indicates that the flat front approximation breaks down after a finite time due to the crude simplifications of the model in this limit. Having nonzero $\Lambda$, $\alpha$ and $\beta$ enhance the validity of the flat front approximation avoiding this finite-time singularity, as we will show in the following sections.

Fig.~\ref{fig:fig1}(c) shows the perturbation amplitude $L_1$ as a function of time $t$, for different values of the wavenumber $k$ indicated in the legend, obtained from the linear stability analysis (eqn~\eqref{eq:eqL1}) (solid lines), and from the two-dimensional numerical simulations of the complete equations of motion (filled circles). Initially, the growth of $L_1$ is exponential for all values of $k$. In particular, for small values of $k$, this initial exponential growth is observed only at very short times, whereas when the value of $k$ increases, the exponential growth lasts longer and the instantaneous growth rate $\omega_{\textrm{inst}} = \textrm{d}(\ln(L_1))/\textrm{d}t = (\textrm{d}L_1/\textrm{d}t)/L_1$ varies very slightly with respect to the initial plateau. We have denoted this initial exponential growth rate as $\omega_{\textrm{s}}(k)$, which is shown in the inset of Fig.~\ref{fig:fig1}(c) as a function of $k$. In this limit, all the perturbation wavelengths are unstable, as already shown in ref.~\cite{Alert2019}. The function $\omega_{\textrm{s}}(k)$ exhibits a maximum value at finite $k$, namely $k \simeq 1.48$, which was also reported in ref.~\cite{Alert2019}, where the authors analyse the growth rate $\omega(k,t)$ for a frozen value of $L_0$. Beyond this value, $\omega_s$ decreases as $k^{-1}$, whereas for $k = 0$, $\omega = 1/3$. 

To assess the dynamical stability of the spreading front, it is important to analyse the complete time evolution of the perturbation amplitude $L_1$, not just its short-time behaviour. Indeed, although the amplitude grows initially faster at relatively large values of $k$, $L_1$ eventually grows faster at small values of $k$, as shown by Fig.~\ref{fig:fig1}(c), thereby evidencing the significance of analysing the whole time evolution. In particular, for $k \lesssim 0.05$, $L_1(t)$ becomes independent of $k$, except very close to $t = 6$, where $L_1$ still grows faster as the value of $k$ decreases. It is also important to emphasise that, even if $L_1 \sim O(1)$, the linear stability analysis still accurately describes the evolution of the migrating monolayer, which means that the perturbation remains spatially harmonic during most of the spreading process.

\subsection{The role of edge polarisation}\label{subsec:case_2}
Here we discuss the effect of edge polarisation and weak bulk polarisation, by considering finite values of the dimensionless parameter $\Lambda$, while still keeping $\alpha = \beta = 0$ in eqn~\eqref{eq:u_flat_complete}. We will show that a nonzero $\Lambda$ delays the breakdown of the flat front approximation, thereby enhancing the applicability of the linear stability analysis.

\begin{figure*}[ht!]
    \centering
    \includegraphics[width=0.9\textwidth]{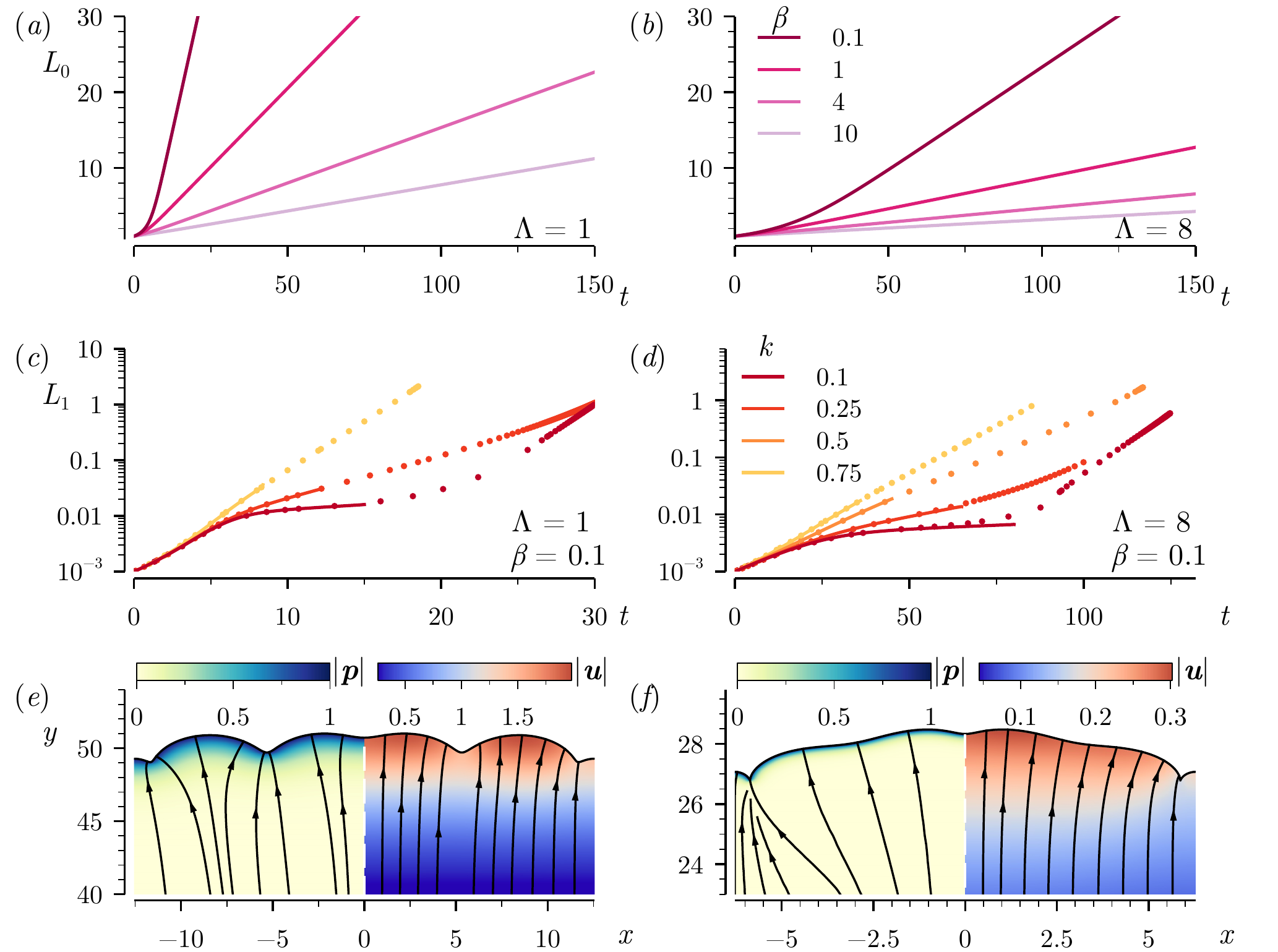}
    \caption{(a,b) Base flow interface $L_0$ for (a) $\Lambda$= 1 and (b) $\Lambda$= 8 as a function of time $t$ for different values of $\beta$ indicated in the legends. (c,d) Perturbation amplitude $L_1$ as a function of time $t$ for different values of the wavenumber $k$ indicated in the legends, for $\beta = 0.1$ and $\Lambda = 1$, in (c) and for $\beta=0.1$ and $\Lambda = 8$ in (d). Results from linear theory and from numerical simulations are indicated by solid lines and filled circles, respectively. (e,f) Snapshot of the monolayer edge in the nonlinear regime showing $|\boldm{u}|$ and $|\boldm{p}|$, for $\beta = 0.1$, (a) $\Lambda = 1$, $k = 0.25$, $t = 31.613$, (b) $\Lambda = 8$, $k = 0.5$, $t = 116.960$.}
    \label{fig:fig4}
\end{figure*}


Although the details become more cumbersome, the linear stability analysis is performed as in the previous section. In the present case, we analyse the effects of cell polarisation through the monolayer at $\Lambda=1$ and $\Lambda=8$. These values are motivated by experimental measurements~\citep{blanch2017effective,perez2019active}. When $\Lambda = 1$, the cell polarisation decays with a characteristic length comparable to $\ell$. However, when $\Lambda = 8$, only cells located close to the edge of the monolayer are strongly polarised, whereas cells at the bulk remain weakly polarised. Fig.~\ref{fig:fig2}(a) shows the time evolution of $L_0$ for different values of $\Lambda$, evidencing that the monolayer spreading velocity is larger as the value of $\Lambda$ decreases, since the polarisation field $\boldm{p}$ does not rapidly decay close to the interface. In particular, $L_0$ grows exponentially with time, with an $e$-fold time that increases with $\Lambda$, namely $L_0 \sim \exp[t/(2 \Lambda)]$~\citep{Alert2019}.

Additionally, Figs.~\ref{fig:fig2}(b,c) show the perturbation amplitude $L_1$ for $\Lambda = 1$ and $\Lambda = 8$, respectively, and different values of the wavenumber $k$ indicated in the legends. In particular, solid lines are obtained from linear theory and filled circles corresponds with the results obtained from the time-dependent numerical simulations. Here we also observe that, for finite values of $\Lambda$, the perturbation remains spatially harmonic even when $L_1$ is of order unity, yielding and excellent agreement between linear theory and numerical simulations. Moreover, similarly to the dynamics of the base-flow solution, the perturbation amplitude $L_1$ also grows faster as the value of $\Lambda$ decreases. To analyse the transient dynamics of $L_1$, Figs.~\ref{fig:fig2}(d,e) show the instantaneous growth rate $\omega_{\textrm{inst}}$ as a function of time $t$. For finite values of $\Lambda$ and depending on the value of $k$, $L_1$ experiences several crossovers between different exponential regimes, which does not occur in the limiting case of $\Lambda = 0$ due to the finite-time singularity. Thus, as in Section~\ref{subsec:case_1}, we define a short-time and also a long-time exponential growth rate, denoted by $\omega_{\textrm{S}}$ and $\omega_{\textrm{L}}$, respectively. These growth rates are shown in Fig.~\ref{fig:fig3}(a) as functions of $k$, for the different values of $\Lambda$ indicated in the legend. The long-time growth rate $\omega_{\textrm{L}}$ is well defined only when $\omega_{\textrm{inst}}$ reaches a plateau for $t \gg 1$, and it is obtained numerically by imposing the criterion $\textrm{d}\omega_{\textrm{inst}}/\textrm{d}t < 10^{-2}$.

\begin{figure*}[ht!]
    \centering
    \includegraphics[width=\textwidth]{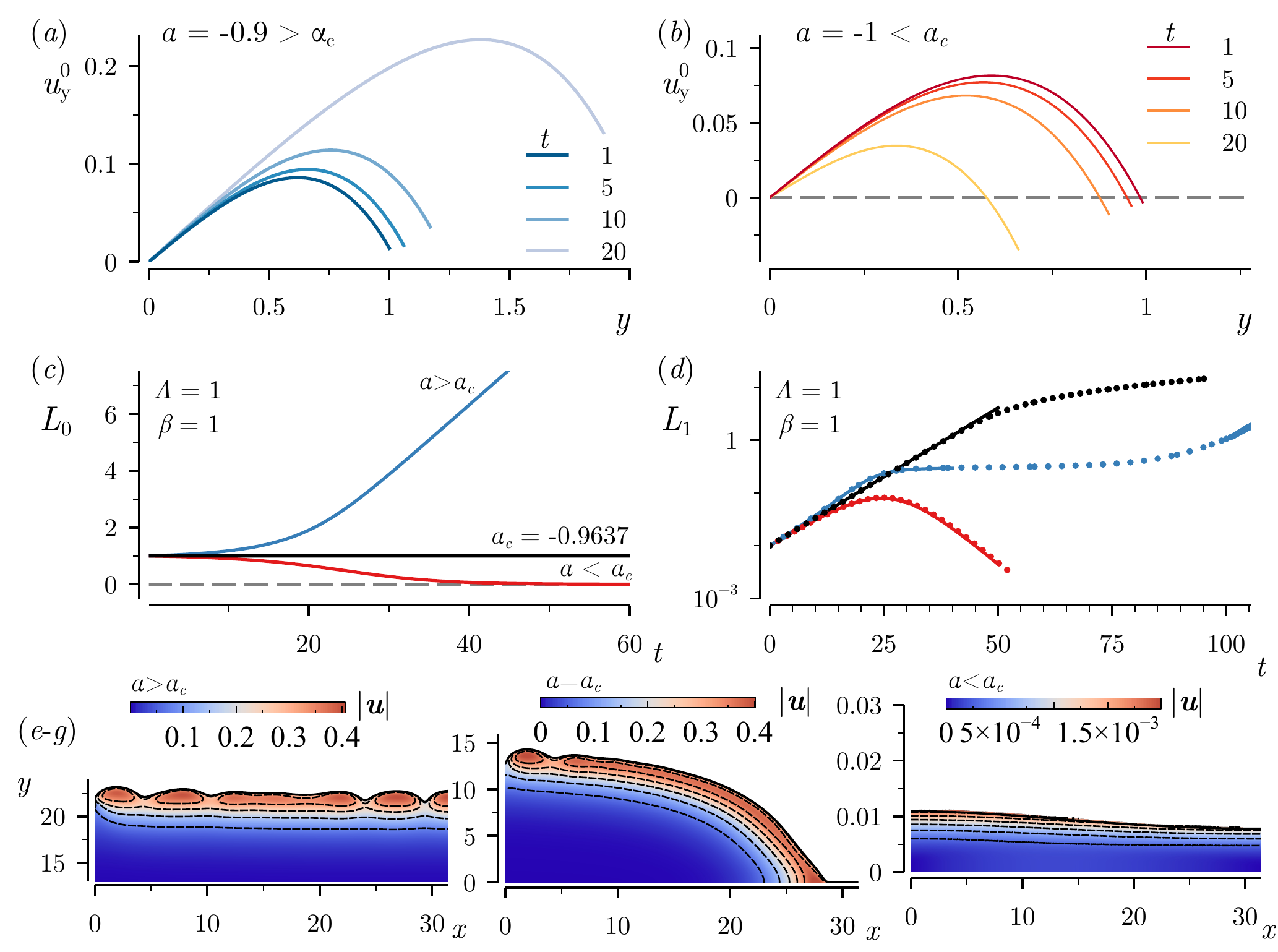}
    \caption{(a,b) Flat-front velocity field $u_y^0(y)$ at different times indicated in the legend, for $\Lambda = 1$, $\beta = 1$, (a) $\alpha = -0.9 > \alpha_{\textrm{c}}$, and (b) $\alpha = -1 < \alpha_{\textrm{c}}$. (c) Flat-front edge time evolution $L_0(t)$ for the same values of $\Lambda$ and $\beta$, $k = 0.1$, and the three cases considered for $\alpha$, namely expanding $\alpha = -0.9$, contracting $\alpha = -1$, and marginally quiescent state $\alpha \simeq -0.96$. (d) Time evolution of the perturbation amplitude $L_1(t)$ for the same values as in (c). Panels (e-g) show snapshots of the three cases considered in (c,d)}
    \label{fig:fig5}
\end{figure*}

Both, short-time (solid lines) and long-time (dashed lines) growth rates decrease monotonically as the value of $\Lambda$ increases for every value of $k$. The most-amplified short-time growth rate occurs at finite $k$ for every value of $\Lambda$, as already pointed out in the limiting case $\Lambda = 0$, discussed in Section~\ref{subsec:case_1}~\citep{Alert2019}. Figs.~\ref{fig:fig3}(b,c) show the maximum short-time growth rate $\omega_{\textrm{S},\textrm{m}}$ and the corresponding most-amplified wavenumber $k_{\textrm{S},\textrm{m}}$ as functions of $\Lambda$. While the maximum short-time growth rate decreases with $\Lambda$, the most-amplified wavenumber increases monotonically with $\Lambda$. However, at long time, the most-amplified growth rate occurs at $k \ll 1$, with a maximum value that is larger than the maximum short-time growth rate for all values of $\Lambda$. At large values of $k$, namely $k \gtrsim 5-10$, short-time and long-time growth rates coincide, since the growth of $L_1$ remains nearly exponential during the whole time evolution.

\subsection{The role of substrate friction}\label{subsec:case_3}

In this section, we analyse the effect of cell-substrate friction in the limit where contractile stresses are negligible, $\alpha = 0$. Eqn~\eqref{eq:p_flat_complete} gives the base-flow polarity and eqn~\eqref{eq:u_flat_complete} with $\alpha=0$ produces the velocity field. The latter
 is a positive function for $y>0$ (and positive values of the dimensionless parameters) whose maximum value is located at the edge of the monolayer as in the previous limiting cases. 

The temporal evolution of $L_0$ is shown in Figs.~\ref{fig:fig4}(a,b) for different values of $\beta$ indicated in the legend, and (a) $\Lambda = 1$, (b) $\Lambda = 8$. After a short transient, $L_0$ grows linearly with time, with constant velocity $V = 1/(\beta + \sqrt{2 \beta} \Lambda)$, in agreement with the experimental observations in ref.~\cite{blanch2017effective}. As expected, $L_0$ is slowed down by the action of cell-substrate friction.

The linear stability analysis is performed in the same manner as in the previous cases. Figs.~\ref{fig:fig4}(c,d) display the time-dependent evolution of the perturbation amplitude $L_1$ for the different values of the wavenumber $k$ indicated in the legend, with (c) $\Lambda = 1$, (d) $\Lambda = 8$, and $\beta = 0.1$. The displayed results are extracted from linear theory (solid lines), and from numerical simulations (filled circles). By comparing these results with those shown in Figs.~\ref{fig:fig3}(b,c) for $\beta = 0$, we observe that substrate friction has a stabilising effect on $L_1$. Additionally, at short time, the most amplified wavenumber is shifted towards larger values, as already pointed out in ref~\cite{Alert2019}. More interestingly, for small values of $k$, e.g., $k=0.1$, the perturbation amplitude almost reaches a plateau within the linear regime, as shown in Figs.~\ref{fig:fig4} (b,c). However, as time evolves, the linear stability analysis cannot predict the growth of $L_1$, even though it is far from being of order unity. After the plateau, $L_1$ increases with a larger slope than its short-time growth rate indicating the onset of secondary instabilities. 

\begin{figure*}[ht!]
    \centering
    \includegraphics[width=0.95\textwidth]{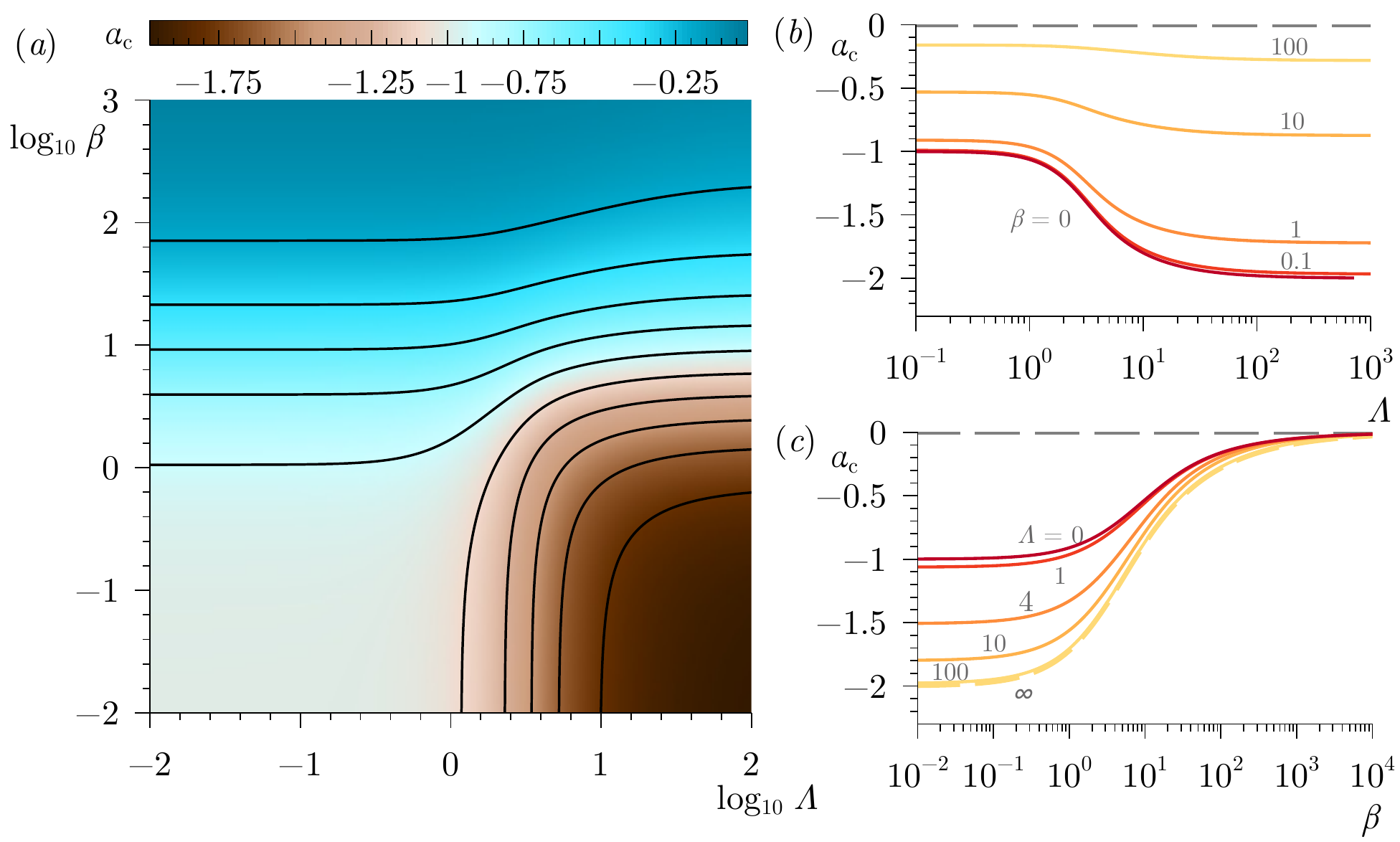}
    \caption{(a) Contour plot of the critical contractile number $\alpha_{\textrm{c}}$ as a function of $\Lambda$ and $\beta$. Panels (b,c) show $\alpha_{\textrm{c}}$ as a function of $\Lambda$ and $\beta$ for different constant values of $\beta$ and $\Lambda$, respectively.}
    \label{fig:fig6}
\end{figure*}

Figs.~\ref{fig:fig4}(e,f) display two snapshots of the spreading monolayer for the same values of $\Lambda$ and $\beta$ as in panels (c,d), at the last computed time, namely (a) $t = 31.613$ with $k = 0.25$, and (b) $t = 116.960$ with $k = 0.5$. These snapshots evidence that cell-substrate friction is responsible for producing a secondary fingering instability in the nonlinear regime. In the absence of friction, the initial perturbation grows according to the linear analysis, i.e. non-trivially in time and remaining spatially harmonic during the whole time evolution, as shown in Figs.~\ref{fig:fig1}(c) and~\ref{fig:fig2}(b,c). The secondary instability resembles the tip-splitting phenomenon in the classical hydrodynamic Saffman-Taylor instability, which occurs for small values of the surface tension coefficient between the two fluids. It is also interesting to point out that, after the initially harmonic shape is destabilised into more finger-like structures, the fluid interface can eventually develop Eden-like structures~\citep{eden1961two} when $\Lambda = 1$ in panel (e), and even more pronounced sulci when $\Lambda = 8$ in panel (f). Although these interfacial patterns have not been observed in spreading epithelial monolayers, they arise in several biological processes, namely in morphogenesis~\citep{hohlfeld2011unfolding,hohlfeld2012scale,shyer2013villification,tallinen2013surface,tallinen2014gyrification,tallinen2016growth}, or during the growth of surface-adhered bacterial communities~\citep{ben1992adaptive,ben1994generic,shapiro1995significances,kearns2010field,ben2012bacterial,farrell2013mechanically}. Indeed, active continuum frameworks similar to the one used in the present work have been used to describe these phenomena, thereby our results are not strictly restricted to spreading epithelial monolayers.

These results suggest that, although cell-substrate friction is able to trigger complex nonlinear edge patterns, contractile forces are necessary to obtain the patterns observed in Fig.~\ref{fig:fig0}. The following section contains a detailed analysis of the role of contractile stresses in the fingering instability.


\subsection{The role of contractility}\label{subsec:case_4}

We now focus our attention to the effect of contractile stresses, that is $\alpha < 0$, on the monolayer spreading dynamics. To this end, we first explore its effect on the baseflow velocity field corresponding with the flat-front solution, defined previously in eqn~\eqref{eq:u_flat_complete}.
\begin{figure*}[ht!]
    \centering
    \includegraphics[width=0.95\textwidth]{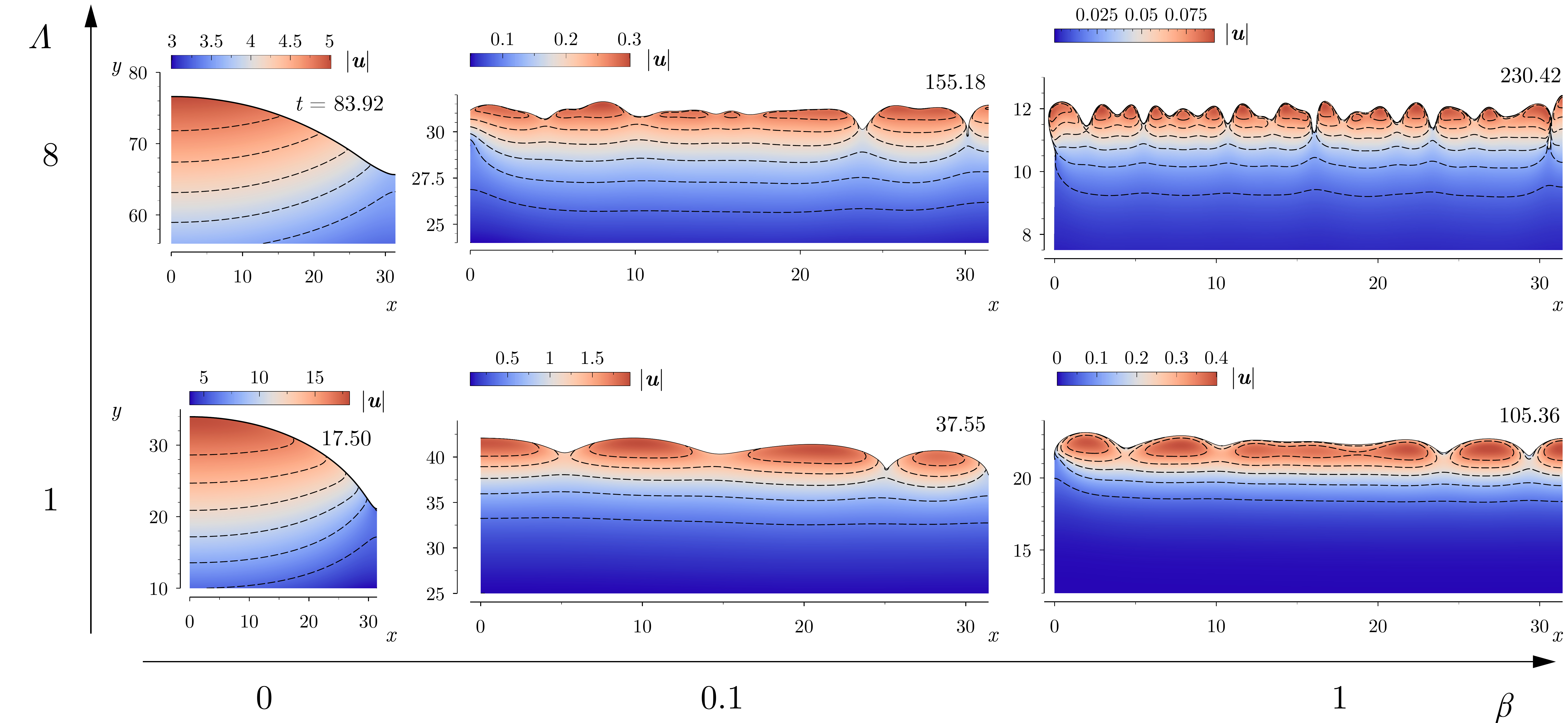}
    \caption{Snapshots of the spreading monolayer in the parameter space ($\Lambda$,$\beta$), for $k = 0.1$, $\alpha = -1$ in the top row ($\Lambda = 8$), and $\alpha = -0.9$ in the bottom row ($\Lambda = 1$). Times are indicated in the labels.}
    \label{fig:fig7}
\end{figure*}
Due to contractility, eqn~\eqref{eq:u_flat_complete} exhibits a maximum value within the bulk, as opposed to the previous limiting cases. In particular, the stream-wise position of maximum velocity depends on the characteristic polarisation length, that is $y_{\max} \sim \Lambda^{-1}$. For large values of $\Lambda$, the maximum velocity is located close to the edge since the monolayer is weakly polarised in the bulk. However, for order-unity and small values of $\Lambda$, the maximum velocity is approximately located between the edge and the middle line. This can be observed in Figs.~\ref{fig:fig5}(a,b), which show $u_y^0(y)$ at different times for $\Lambda = 1$, $\beta = 1$, and (a) $\alpha = -0.9$, (b) $\alpha = -1$. Additionally, Figs.~\ref{fig:fig5}(a,b) show that, due to contractile stresses, the flat-front velocity at the edge $\boldm{u}_0(L_0)$ can be negative depending on the values of the dimensionless parameters. Indeed, we can define a critical contractile number, $\alpha_{\textrm{c}} < 0$, at which $u_y^0(L_0)=0$, meaning that $L_0$ remains quiescent at its initial value:
\begin{equation}
\alpha_{\textrm{c}} = \frac{2(\beta - 8 \Lambda^2) \left[ \sqrt{\beta} - \sqrt{2} \Lambda \coth (\Lambda) \tanh (\sqrt{\beta/2})\right]}{(\beta - 2 \Lambda^2) [\sqrt{2} (\beta + 4 \Lambda^2 \csch(\Lambda)^2)\tanh(\sqrt{\beta/2}) - 4 \sqrt{\beta}\Lambda \coth(\Lambda)]}.
\label{eq:criticalalpha}
\end{equation}
The critical value $\alpha_c(\beta,\Lambda)$  depends on the cell-substrate friction $\beta$ and the cell polarisation $\Lambda$, and it characterises the active wetting-dewetting transition~\citep{douezan2012dewetting,douezan2012wetting,alert2018role,perez2019active}. For $\alpha < \alpha_{\textrm{c}}$, the monolayer contracts with time (dewetting), while for $\alpha > \alpha_{\textrm{c}}$ the monolayer expands (wetting). Fig.~\ref{fig:fig6} shows a colour plot of $\alpha_{\textrm{c}}$ as a function of $\beta$ and $\Lambda$ in (a), and $\alpha_c$ at constant values of $\beta$ and $\Lambda$ in (b,c), respectively. In particular, the absolute value of the critical contractility $|\alpha_{\textrm{c}}|$ decreases monotonically as friction increases for every value of $\Lambda$, meaning that a smaller contractile stress is needed for the monolayer to retract as friction becomes dominant. Indeed, $\alpha_{\textrm{c}} \to 0$ as $\beta \to \infty$. Moreover, $|\alpha_{\textrm{c}}|$ increases monotonically with $\Lambda$ for every finite value of $\beta$, which means that larger contractile stresses are needed for the monolayer to retract when it is weakly polarised within the bulk.

\begin{figure*}[ht!]
    \centering
    \includegraphics[width=0.95\textwidth]{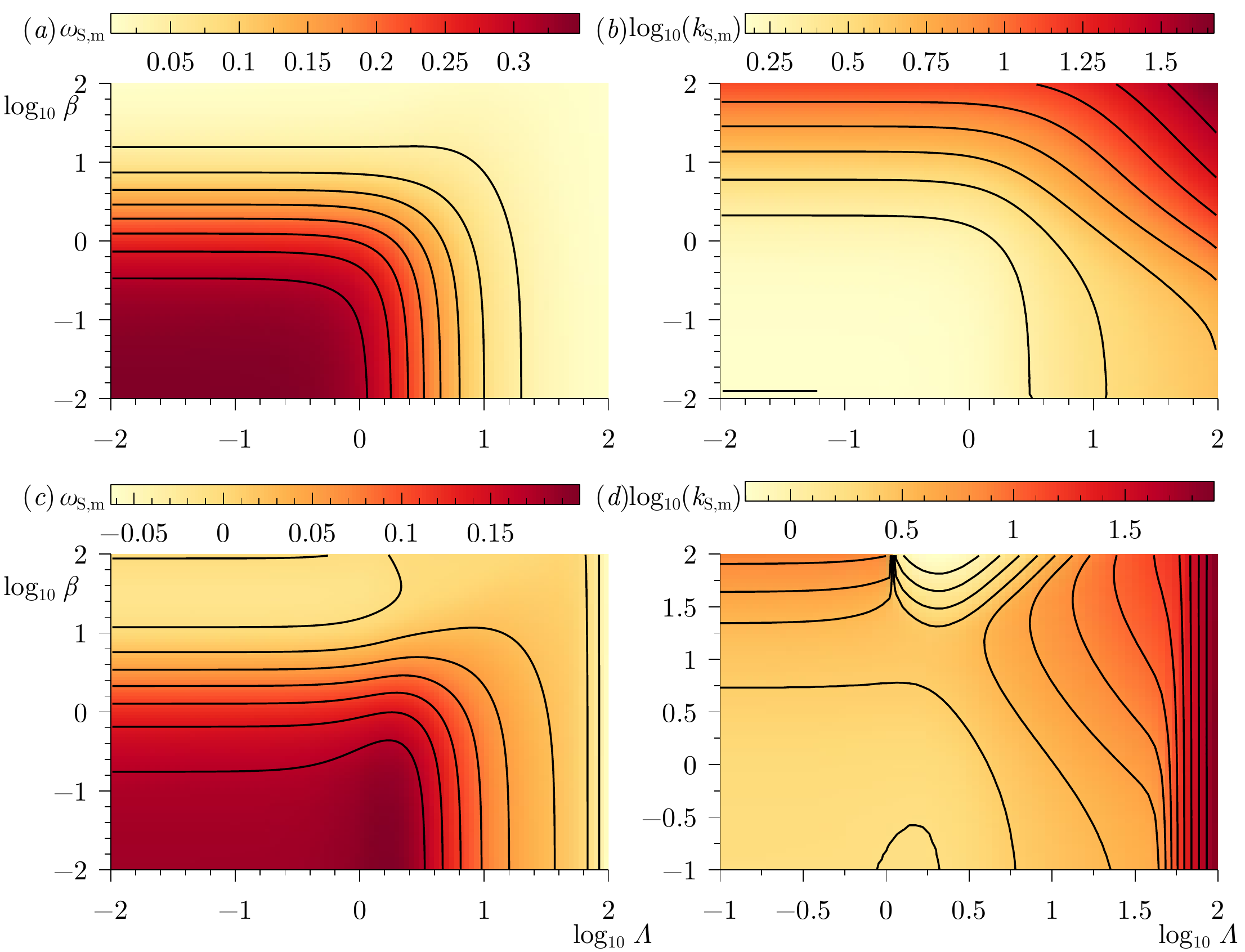}
    \caption{Short-time maximum growth rate $\omega_{\textrm{S},\textrm{m}}$ and the corresponding most amplified wavenumber $k_{\textrm{S},\textrm{m}}$, as functions of $\Lambda$ and $\beta$, for (a,b) $\alpha = 0$, and (c,d) $\alpha = -1$.}
    \label{fig:fig8}
\end{figure*}

The critical transition obtained in the flat-front solution is reflected in Figs.~\ref{fig:fig5}(c) and (e-g), where we show $L_0$ as a function of time $t$ for the expanding ($\alpha = -0.9$), contracting ($\alpha = -1$), and steady-state ($\alpha = \alpha_{\textrm{c}} \simeq -0.964$) cases, and three corresponding snapshots displaying $|\boldm{u}|$ at the last computed time, for $\Lambda = \beta = 1$, and $k = 0.1$. Additionally, Fig.~\ref{fig:fig5}(d) shows the perturbation amplitude $L_1$ as a function of time $t$ for the three different cases considered in panel (c), obtained from linear theory (solid lines), i.e eqn~\eqref{eq:L1_sol}, and from numerical simulations (filled circles), displaying a perfect agreement in the linear regime. In the critical case, $\alpha = \alpha_{\textrm{c}}$, the monolayer edge grows exponentially in the linear regime since initially $\boldm{u}(L,t = 0) = \boldm{0}$. As time evolves, the monolayer adopts a rounded shape without developing a significant number of finger-like protrusions at the edge, as displayed in the snapshot in panel (f). When $\alpha > \alpha_{\textrm{c}}$, the monolayer expands and eventually develops a secondary fingering instability with a smaller characteristic wavelength, similarly to the case with negligible contractile stresses discussed in Sec.~\ref{sec:sec3}. Under dewetting conditions, $\alpha < \alpha_{\textrm{c}}$, the monolayer retracts, and although the perturbation amplitude initially grows, it decays at sufficiently long time, thus evidencing the importance of assessing the time-dependent stability of the moving front. Indeed, we have found that the flow generated by the retracting front is able to stabilise all wavenumbers. This can be explained alluding to the local friction at the edge, which increases as the retraction velocity becomes larger, as shown in Fig.~\ref{fig:fig5}(b). Examining the dependence of $\omega(k,t)$ on $\beta$, we find that, when friction dominates, i.e. for a large value of $\beta$, the maximum growth rate $\omega_{\textrm{m}}$ becomes negative, thus the whole amplification curve becomes negative. This is depicted in Fig.~\ref{fig:fig8}, which shows the short-time maximum growth rate $\omega_{\textrm{S},\textrm{m}}$ and the corresponding most amplified wavenumber $k_{\textrm{S},\textrm{m}}$ as functions of $\Lambda$ and $\beta$ for (a,b) $\alpha = 0$ and (c,d) $\alpha = -1$.

Finally, Fig.~\ref{fig:fig7} displays different snapshots at long time in the parameter space ($\Lambda$,$\beta$) for $\alpha \gtrsim \alpha_{\textrm{c}}$ and $k = 0.1$. For negligible friction, $\beta=0$, these morphologies evidence that the shape of the front remains harmonic for long times, even when contractile stresses play a role. Thus, the main mechanism responsible for tip-splitting phenomena and secondary fingering instabilities is cell-substrate friction. Furthermore, for increasing values of $\Lambda$ and $\beta$, the characteristic wavelength of the secondary instability decreases and the finger-like structures at the edge become more complex. This can be explained in terms of the results of Fig.~\ref{fig:fig8}. This figure shows that the most amplified wavenumber increases with increasing $\beta$ and $\Lambda$. Hence, the resulting small wavelength instability could explain continuing fragmentation of the tissue edge associated with the secondary instability observed in the complete numerical simulations, since the local cell-substrate friction at the edge, proportional to $\boldm{u}$, increases monotonically as the monolayer spreads.

\section{Concluding remarks and future prospects}\label{sec:conclusions}
Motivated by observations of interfacial instabilities in biological contexts during \textit{in vivo} and \textit{in vitro} processes, here we have analysed some of the main mechanisms involved in the appearance of finger-like instabilities in migrating epithelial monolayers. To this end, we have modelled the monolayer as a continuum compressible polar fluid and analysed the linear and nonlinear dynamics of the spreading front. By means of linear stability analyses and numerical simulations, we have unravelled the roles of cell polarisation, substrate friction, and contractile stresses. 

Regarding the linear stability of the spreading front, we have shown that it is crucial to analyse the transient dynamics of the perturbation amplitude $L_1$, since, although it initially grows exponentially in time, it eventually develops non-trivial transient dynamics. In particular, in the limit when cell-substrate friction and contractile stresses are negligible, there is finite most unstable wavenumber at short times~\citep{Alert2019}, but at long time, $L_1$ eventually grows faster for monotonically decreasing values of $k$. This is possible since, for small values of $k$, $L_1$ experiences several crossovers between different exponential regimes. Additionally, in this limiting case, the linear stability analysis is able to describe the transient evolution of the spreading monolayer even at large times when $L_1 \sim O(1)$, meaning that the growth of $L_1$ remains linear and spatially harmonic. 

Only when cell-substrate friction is considered, secondary finger-like structures arise in the nonlinear regime, similar to the ones observed in many biological processes. In this scenario, the linear stability analysis is of limited value, as it is quantitatively valid only when the perturbation amplitude remains small, $L_1 \lesssim 10^{-2}$ (one hundredth of the initial monolayer width). Above this value, secondary edge instabilities arise. Additionally, the fingering pattern becomes more complex when cell-substrate friction becomes dominant over viscous forces and the edge of the monolayer is strongly polarised (i.e., polarisation length much smaller than the initial monolayer width). This result of linear stability analysis could potentially explain the onset of  fingering instabilities. It suggests that fingering instabilities appearing in \textit{in vivo} and \textit{in vitro} experiments may be a direct consequence of cell-substrate friction becoming dominant over viscous forces.

Finally, when contractile stresses are taken into account, we have obtained a critical contractility, $\alpha_{\textrm{c}} < 0$, at which the monolayer displays a quiescent polarised steady state, thus characterising the active wetting-dewetting transition~\citep{perez2019active}. Below this critical value, the monolayer contracts (dewetting, contractile stresses dominate active force traction), whereas it expands above the critical contractility (wetting, active force traction dominate contractile stresses). These results are similar to those previously obtained with short-time linear stability theory and observed in experiments on circular tissues, in which cell contractility is tuned by E-cadherin supply~\citep{perez2019active}. Additionally, in the dewetting scenario, our numerical simulations show that, although the monolayer is initially unstable, it eventually becomes stable as the edge retracts, thus evidencing the importance of assessing the time-dependent evolution of the perturbation amplitude.

Although our work shows that cell-substrate friction is crucial to trigger the formation of finger-like structures at the edge of a migrating epithelial monolayer, a thorough comparison with controlled experiments would be necessary to test these theoretical and numerical results. Our continuum model cannot discern cellular scales that are important in developed fingers. This includes observations of faster cells having larger area at the fingers (which may be interpreted as leader cells with different phenotype) or swirl patterns near the tissue edge~\citep{petitjean2010velocity,sepulveda2013}. These effects can be accounted for by using an active vertex model that includes a collective inertia but not leader cells~\cite{bonilla2020tracking}. Within a continuum model similar to the present one, experiments involving cells whose motion is interrupted by a flexible fibre are explained by adding an effective inertia and relaxing the assumption that there is a large separation between the time scales of polarisation and flows at the monolayer~\citep{valencia2020}. Our work could pave the way to ascertaining the role of these effects in a more complete stability analysis of finger formation.

From a theoretical point of view, cell proliferation would likely play a relevant role during cell migration and the formation of finger-like protrusions, thus the effect of incorporating a density field and a density-dependent pressure deserves further investigation. Other natural extensions of our work include the effect of noise, rheology, durotaxis, coupling between flow and cell polarisation, or chemical signaling, both, on the linear and nonlinear dynamics.

\begin{appendix}
\section{Appendix: Active polar fluid for tissue spreading}\label{app:appendix}
This Appendix contains a brief derivation of the continuum model considered in the present work and, previously, in Refs.~\cite{alert2018role,Alert2019,blanch2017effective,alert2020physical}. The continuum model is based on ideas from liquid crystal theory and active-gel physics~\citep{de1993,prost2015active}. In this approach, the coarse-graining scale is assumed to be at the multicellular level, thus the dynamics of the epithelial sheet is described in terms of the average cell polarisation field $\boldm{p}(\tilde{\boldm{x}},\tilde{t})$, and the velocity field $\tilde{\boldm{u}}(\tilde{\boldm{x}},\tilde{t})$. 
\paragraph*{Polarisation field in epithelial monolayers.} In a spreading epithelial monolayer, cells are polarised near the free edge by contact inhibition of locomotion, which forces the cells to migrate towards the free space, whereas they are not polarised far from the edge. Polarisation is established much faster than the flows in the monolayer, which occur on the strain rate time scale. Thus, we may consider an instantaneous relaxation of the polarisation uncoupled from cellular flows. From the biophysical point of view, polarisation occurs due to cell-substrate active traction forces related to forces exerted by the actomyosin cytoskeleton of cells at focal adhesion sites in the extracellular matrix~\citep{alert2020physical}.



The free energy associated to the polarisation field is 
\begin{equation}
    F = \int_A \left[\frac{a}{2} |\boldm{p}|^2 + \frac{K}{2}\bnabla \boldm{p} : \bnabla \boldm{p}  \right] \textrm{d} A,
\end{equation}
where $a>0$ is a restoring coefficient, $K$ is the
Frank elastic constant in the one-constant approximation~\citep{de1993}, and $A$ is the area. The first term in $F$ favours the isotropic non-polarised state, $\boldm{p}=\boldm{0}$, far  from the monolayer free edge. The second term describes the cell-cell polarity interactions. We have neglected higher-order terms and the coupling with the cell density field. The first variation of $F$ with respect to $\boldm{p}$ yields the instantaneous polarisation, 
\begin{equation}\label{eq:dim_p}
L_c^2 \bnabla^2 \boldm{p} = \boldm{p},
\end{equation} 
where $L_{\textrm{c}} =\sqrt{K/a}$ is the nematic length. We impose a homeotropic boundary condition at the monolayer free edge, i.e. the polarity field is anchored perpendicular to the interface. Thus, cells are polarised close to the free edge of the tissue (with characteristic length $L_{\textrm{c}}$), where they are not contact inhibited.  
\paragraph*{Stress balance.} In addition to eqn~\eqref{eq:dim_p}, the balance of forces within the epithelial monolayer is 
\begin{equation}\label{eq:mom}
\bnabla \bcdot \tilde{\boldm{\sigma}} + \tilde{\boldm{f}} = \boldm{0},
\end{equation}
where $\tilde{\boldm{\sigma}}(\tilde{\boldm{x}},\tilde{t})$ is the stress tensor, and  $\tilde{\boldm{f}}(\tilde{\boldm{x}},\tilde{t})$ is the body force due to cell-substrate interactions. Multiplying by the tissue height $h$, $h\tilde{\boldm{\sigma}}(\tilde{\boldm{x}},\tilde{t})$ and $h\tilde{\boldm{f}}(\tilde{\boldm{x}},\tilde{t})$ are the measured monolayer tension and traction stress fields, respectively~\citep{perez2019active}. For simplicity, we assume that the stress tensor contains only viscous and contractile stresses, $\tilde{\boldm{\sigma}} = \mu (\bnabla \tilde{\boldm{u}} + \bnabla \tilde{\boldm{u}}^{\textrm{T}}) - \zeta \boldm{p} \boldm{p}$. We neglect the short time elastic response of the tissue and pressure effects, which, in the absence of cell proliferation, are much smaller than the tensile stress induced by traction forces~\citep{Alert2019}. The body force consists of cell-substrate forces: cell-substrate friction proportional to the velocity field, and the active polar traction force driving the migration, which is proportional to the polarity field, $\tilde{\boldm{f}} = -\xi \tilde{\boldm{u}} + (T/h) \boldm{p}$~\citep{alert2020physical}. Additionally, we have assumed that inertial forces are negligible since flows within epithelial monolayers typically occur at very low Reynolds numbers.


\end{appendix}

\section*{Conflicts of interest}
There are no conflicts to declare.

\section*{Acknowledgements}
CT and LLB acknowledge financial support by the FEDER/Ministerio de Ciencia, Innovaci\'on y Universidades -- Agencia Estatal de Investigaci\'on grant MTM2017-84446-C2-2-R, by the Madrid Government (Comunidad de Madrid-Spain) under the Multiannual Agreement with UC3M in the line of Excellence of University Professors (EPUC3M23), and in the context of the V PRICIT (Regional Programme of Research and Technological Innovation). AM-C acknowledges support from the Human Frontier Science Program (LT000035/2021-C), and from the FEDER/Ministerio de Ciencia, Innovaci\'on y Universidades -- Agencia Estatal de Investigaci\'on through the project DPI2017-88201-C3-3-R, and the Red Nacional para el Desarrollo de la Microflu\'idica, RED2018-102829-T. The authors warmly acknowledge Ricard Alert-Zen\'on for useful comments and insightful advice.

\bibliography{biblio}
\bibliographystyle{rsc}

\end{document}